\documentclass[usenatbib,useAMS]{mnras}
\pdfoutput=1

\usepackage[latin1]{inputenc}  
\usepackage{graphicx}
\usepackage[T1]{fontenc} 
\usepackage{color}
\usepackage{amssymb}
\usepackage{float}
\usepackage{times}
\usepackage{multirow}

\newcommand{\gl}{\textsc{galform}}
\newcommand{\eg}{\textsc{eagle}}
\newcommand{\egdm}{\textsc{eagleDMO}}
\newcommand{\lgl}{\textsc{l-galaxies}}
\newcommand{\subf}{\textsc{subfind}}
\newcommand{\msun}{{\rm M}_{\odot}}
\newcommand{\mb}{M_{\rm Break}}
\newcommand{\mth}{{\rm M}_{200}^{\rm crit}}
\newcommand{\nm}{$\langle N \rangle_{M}$}

\title[Galaxies in \eg~, \gl~and \lgl]{Galaxies in the EAGLE hydrodynamical simulation and in the Durham and Munich semi-analytical models}
\author[Guo et al.]{
\parbox[c]{0.92\textwidth}{
\vspace{-0.8cm}
Quan\,Guo$^{1,2}$\thanks{E-mail: guotsuan@gmail.com (QG)},
Violeta\,Gonzalez-Perez$^{3,2}$\thanks{E-mail: violegp@gmail.com (VGP)},
Qi\,Guo$^{4}$, 
Matthieu\,Schaller$^{2}$, 
Michelle\,Furlong$^{2}$,
Richard\,G.\,Bower$^{2}$,
Shaun\,Cole$^{2}$, 
Robert\,A.\,Crain$^{5}$, 
Carlos\,S.\,Frenk$^{2}$,
John\,C.\,Helly$^{2}$, 
Cedric\,G.\,Lacey$^{2}$,
Claudia\,del\,P.\,Lagos$^{6,7}$,
Peter\,Mitchell$^{2}$, 
Joop\,Schaye$^{8}$,
Tom\,Theuns$^{2}$.}
\vspace*{8pt} 
\\
$^{1}$Leibniz-Institut f\"ur Astrophysik Potsdam, An der Sternwarte 16, D-14482 Potsdam, Germany.\\
$^{2}$Institute for Computational Cosmology, Department of Physics, 
University of Durham, South Road, Durham, DH1 3LE, U.K. \\
$^{3}$Institute of Cosmology and Gravitation, University of Portsmouth, Dennis Sciama Building, Portsmouth PO1 3FX, UK. \\
$^{4}$Key Laboratory for Computational Astrophysics, The Partner Group of Max
Planck Institute for Astrophysics, \\ National Astronomical Observatories, Chinese Academy of Sciences, Beijing, 100012, China.\\
$^{5}$Astrophysics Research Institute, Liverpool John Moores University, 146 Brownlow Hill, Liverpool, L3 5RF.\\
$^{6}$International Centre for Radio Astronomy Research, 
      University of Western Australia, 35 Stirling Highway, 
      Crawley, WA 6009, Australia.\\
$^{7}$Australian Research Council Centre of Excellence for All-sky Astrophysics (CAASTRO), 44 Rosehill Street Redfern, NSW 2016, Australia.\\
$^{8}$Leiden Observatory, Leiden University, P.O. Box 9513, 2300 RA Leiden, the Netherlands.
}

\begin{document}

\label{firstpage}
\maketitle


\begin{abstract}
We compare global predictions from the \eg~hydrodynamical simulation, and two
semi-analytic (SA) models of galaxy formation, \lgl~and \gl. All three models
include the key physical processes for the formation and evolution of galaxies
and their parameters are calibrated against a small number of observables at
$z\approx 0$. The two SA models have been applied to merger trees constructed
from the \eg~dark matter only simulation. We find that at $z\leq 2$, both the
galaxy stellar mass functions for stellar masses $M_*<10^{10.5}\msun$ and the
median specific star formation rates (sSFRs) in the three models agree to better
than $0.4$~dex. The evolution of the sSFR predicted by the three models closely
follows the mass assembly history of dark matter haloes. In both \eg~and
\lgl~there are more central passive galaxies with $M_*<10^{9.5}\msun$ than in
\gl. This difference is related to galaxies that have entered and then left a
larger halo and which are treated as satellites in \gl. In the range $0<z<1$,
the slope of the evolution of the star formation rate density in \eg~is a factor
of $\approx 1.5$ steeper than for the two SA models. The median sizes for
galaxies with $M_*>10^{9.5}\msun$ differ in some instances by an order of
magnitude, while the stellar mass-size relation in \eg~is a factor of $\approx
2$ tighter than for the two SA models.  Our results suggest the need for a
revision of how SA models treat the effect of baryonic self-gravity on the
underlying dark matter. The treatment of gas flows in the models needs to be
revised based on detailed comparison with observations to understand in
particular the evolution of the stellar mass-metallicity relation.


\end{abstract}


\section{Introduction}

The formation and evolution of galaxies in a cosmological context involves a
multitude of physical processes, such as stellar and active galactic nuclei (AGN) feedback, that are hard to constrain directly by observations
\citep[e.g.][]{somerville2014}. Many of these poorly constrained processes are pivotal
for addressing fundamental questions concerning the growth of structure in the Universe. 

In the current paradigm, the ${\rm \Lambda CDM}$ cosmological model, galaxies are formed
in the potential wells generated by the gravity of the underlying dark matter
distribution \citep[e.g.][]{white78,blumenthal86,white91}, which is
assumed to evolve through gravitational interactions
\citep{peebles80,Davis1985}. There have been
two main approaches to understanding the formation and evolution of galaxies: semi-analytical (SA) models and hydrodynamical simulations. 

The fundamental difference between the two approaches is that hydrodynamical simulations simultaneously solve the equations of gravity and
hydrodynamics for dark matter, gas and stars
\citep[recent examples include][]{Oppenheimer2010,Puchwein2013,
  dubois14,Okamoto2014,vogel14,kha14,steinborn15}, while SA models follow the evolution of the gas partitioned by the dark matter halo it occupies \citep[recent examples include][]{qi13,lee14,lu2014,gp14,ruiz15,henriques14,lacey15}. Dark matter haloes are defined as $\approx 200$ times overdense dark matter structures that are gravitationally bound. The evolution of dark matter haloes can be described by merger trees, which are hierarchical structures recording the haloes mass growth and merger history over cosmic time. The halo merger trees needed by SA models are generated from either N-body dark matter simulations or using Monte Carlo techniques \citep[e.g.][]{parkinson08}. 

Another important difference between the two methods is that, while in hydrodynamical simulations no a priori assumptions need to be made about the properties of galaxies, in SA models, baryons are divided into discrete components: gas halo, gas disc, stellar disc,
gas bulge, stellar bulge. Each component has idealised spatial, thermal and velocity
profiles. SA models work by integrating differential equations that describe how mass, energy, angular momentum and metals are exchanged between these different components. These equations encapsulate the physical processes that are thought to be the most relevant for
the formation and evolution of galaxies \citep[see][for reviews of
SA models]{bau06,benson10}. 

Both SA models and hydrodynamical
simulations require subgrid prescriptions. SA models use physical models dependent on global galaxy properties, for example in supernova (SN) feedback the wind speed can depend on the disc circular velocity. Hydrodynamical simulations attempt to model the physical processes in more detail than SA models, since in these simulations the gas and star particles are followed explicitly. For example, the radiative cooling rate in hydrodynamical simulations is, in principle, determined by
atomic physics (as it is in SA models). Nevertheless, subgrid models on the scale of the interstellar medium (ISM) structure, are necessary to account for the limited numerical
resolution and physics in hydrodynamical simulations. Moreover, in both hydrodynamical simulations and SA models, most subgrid models are a mixture of physical and empirical
ingredients which require calibration.  

Hydrodynamical simulations and SA models are complementary, with their own
advantages and disadvantages. The first capture the complexity of gas and
stellar dynamics in a way that SA models cannot, which is invaluable when
studying, for example, the details of gas accretion onto galaxies. They are,
however, limited by the extraordinary computational cost of having to resolve
very small scales over cosmological volumes. Therefore, the simulated volumes
have been modest \citep[e.g.][]{genel12,zolotov12,marinacci14,hopkins14},
although hydrodynamical simulations of cosmological volumes have started to
become available \citep{Dubois2014,Hirschmann2014,Khandai2015}, and multiscale
approaches are needed to account for processes that cannot be directly resolved
\citep[e.g.][]{vogel14,crain15}. SA models, on the other hand, are more flexible
and require a relatively modest computational cost. Thus, these models can be
run over large cosmological volumes and can also be used to explore extensively
the parameter space in a statistically significant way employing techniques such
as Monte Carlo Markov chains \cite[e.g.][]{henriques09,lu11,benson14}, particle
swarm \citep{ruiz15} and emulator techniques \citep{Bower2010}. Although
the flexibility of SA models is achieved at the cost of approximations that
might be more inaccurate than those introduced in hydrodynamical simulation,
these models predict global galactic properties that are qualitatively similar
\citep[e.g.][]{DeLucia2010,Fontanot2011,lu2014,Knebe2015}.

Previous comparisons between hydrodynamical simulations and SA models have focused on
either comparing a handful of objects \citep[e.g.][]{str10,hir12} or on a single
aspect of the physics, for example gas cooling
\citep{ben01,hel03,yos02,Benson2011,Lu2011}. In some of the latter cases,
stripped-down versions of SA models are used, in which all the other physical
processes relevant for galaxy formation apart from that under scrutiny are
switched off \citep[e.g.][]{cat07,sar10,nei12,mon14}. In the case of gas cooling
studies, this usually implies turning off star formation and thus ignoring the
gas flowing back into the halo due to feedback. Thus, such studies are limited
and thus it is important to also compare complete and realistic models of galaxy
formation and evolution as those presented here. 

Recently, the first cosmological hydrodynamical simulations that reproduce some of the fundamental observations of the local galaxy population have been published \citep[e.g.][]{vogel14,schaye15}. This has been achieved by
modelling and calibrating the subgrid physical processes that shape the gas
cooling, the star and black hole (BH) formation, the metal enrichment and the
stellar and AGN feedback. Furthermore, the parameters of the subgrid models have
been constrained by comparing the hydrodynamical simulation results against
observations, in a similar way as has traditionally been done for SA models.
The new hydrodynamical simulations have been run in cosmological volumes,
although these remain a factor up to $\approx 400$ smaller than that used for
the Millennium simulation \citep{Springel2005}, which is the basis of several SA
models \citep[e.g.][]{bow06,dlb07}. 

Given these recent advances in hydrodynamical simulations and the overlap in physical processes modelled, it is now an appropriate time to examine thoroughly the similarities and differences of galaxy samples produced by different complete galaxy formation models. This comparison will allow us to explore the impact of variations in the subgrid implementation across the models. Such a comparison will not only help in exploring better parameterizations of subgrid physics, but it will lay the foundations for comparing galaxies in greater
detail on smaller scales, where the advantages of hydrodynamical simulations are
most important. 

Recently, \citet{somerville2014} compared results from published hydrodynamical simulations and SA models, finding remarkable agreement between the global properties of model galaxies, such as the galaxy stellar mass (GSMF) function and the stellar mass-star formation rate relations. In this work we expand on that study, by comparing a hydrodynamical simulation
with two semi-analytical models built upon merger trees extracted from dark
matter only realisations of the same initial conditions as the
hydrodynamical simulation. 

\eg~is one of the first cosmological hydrodynamical simulation to model galaxy populations whose basic properties are broadly consistent with observations \citep{schaye15,crain15,fur14,Furlong2015,Trayford2015,lagos15}. This  level of agreement with observations has been achieved by calibrating the free parameters of the subgrid physics modelling to match the observed local GSMF and stellar mass-size relations. A similar level of agreement with observations was previously only achieved by SA models and ad hoc empirical models constructed with that specific purpose \citep[e.g.][]{favole15}. In this paper we make the
first detailed comparison between the results from this state-of-the-art
cosmological hydrodynamical simulation and two SA models, \gl~and \lgl. 

In order to make a fair
comparison, while focusing on the modelling of the physical processes relevant for galaxy evolution, we have generated dark matter merger trees from the \eg~dark matter-only (\egdm) simulation, populating the haloes with galaxies using SA models. Some small changes have been introduced in the \gl~model in order to use the same initial mass function (IMF) as in \eg~and \lgl, and also to obtain a similar level of agreement with the observed passive fraction of galaxies at $z=0$. The model parameters in \lgl~and \gl~are separately calibrated against observations according
to their own criteria, namely the $z=0$ GSMF and luminosity functions respectively (see
\S~\ref{sec:calibration} for details).  The models are therefore \textit{not}
explicitly calibrated to match each other.  

Based on the output galaxy catalogues, we
compare the predictions of several global properties of model galaxies. In this
work we are interested in exploring the similarities and differences in the
model galaxy population with the aim of probing the physics included in the
model. The three models used in this paper describe the same set of key physical
processes, thought to be relevant to galaxy formation and evolution, however,
the details of the implementation can be quite different among the models. Thus,
it is interesting to compare global galactic properties from different modelling
approaches in order to understand the origin of the largest discrepancies. This
work is also meant as the fundation for a future comparisson between individual
objects (Mitchell et al., in prep). The aim of this paper is to determine the
similarities and differences between the three models and understand the
physical origin of the latter. This work does not aim to establish which model
performs better compared to a certain set of observable. Thus, we include only
very limited observational data just as a reference. Moreover, no attempt is
made in this work to discuss how the models compare with the presented
observations as this will detract the text from our main purpose. Detailed
comparisons of \eg, \lgl~and\gl with observations can be found in other studies
\citep{qi11,gp14,schaye15, fur14,Furlong2015, lagos15,henriques14,lacey15}.

The remainder of this paper is organised as follows. In Section
\ref{sec:models}, we briefly introduce the \eg~simulation and the two
semi-analytical models, \gl~and \lgl. In \S\ref{sec:sm} we compare the
GSMFs, the stellar mass-halo mass relations for central galaxies and the halo
occupation distributions of the models. The star-forming sequence is defined in
\S\ref{sec:sf}, where properties of galaxies separated into star-forming and
passive populations are explored. The mass-metallicity and mass-size relations
are discussed in \S\ref{sec:Z} and \S\ref{sec:r50}, respectively. The Appendix
contains a discussion on the definition of the halo masses in the models. The conclusions of
this work are presented in \S\ref{sec:conclusions}.   

\section{Models of galaxy formation}\label{sec:models}
For this work we compare global galaxy properties predicted by the
hydrodynamical simulation \eg~and two different semi-analytical (SA) models of
galaxy formation, \gl~and \lgl, built on the halo merger trees from the \eg~dark
matter-only (\egdm) simulation. The \egdm~and main \eg~simulations model the same volume, with initial conditions produced using the same phases except that the volume of \egdm~is sampled with $1504^3$ dark matter particles, while \eg~is sampled with $1504^3$ dark matter and $1504^3$ baryonic particles. The properties of the simulations are summarised in Table~\ref{tbl:eagle}. The simulations assume the $\Lambda$CDM best-fitting cosmological parameters from the \cite{planck13} data given in Table~\ref{tbl:cosmo}. The initial conditions were generated using second order Lagrangian perturbation theory with the method of \cite{Jenkins2010} and evolve from $z=127$.

\begin{table} 
  \caption{Parameters characterising both the \egdm~and main \eg~simulations. From
top-to-bottom the rows show: comoving box size; number of dark matter particles
(there is initially an equal number of baryonic particles); initial gas particle mass for the main \eg~simulation; dark matter particle mass for the \egdm~and the \eg~simulations; comoving gravitational softening length; maximum physical softening length.
  }
\begin{center}
\begin{tabular}{|l|r|}
\hline
Property & \egdm, \eg \phantom{00000000} \\
\hline
   $L$ (comoving Mpc)             &   100\phantom{000000000000}  \\
   $N$             &    $1504^3$\phantom{0000000000} \\
   $m_{\rm g}$     &      $-,1.81\times10^{6}~\msun$ \\ 
   $m_{\rm dm}$    &    $1.15\times10^{7}~\msun,9.70\times10^{6}~\msun$ \\
   $\epsilon$~(comoving kpc), $z>2.8$  &    2.66\phantom{000000000000}   \\
   $\epsilon$~(proper kpc), $z<2.8$    &    0.70\phantom{000000000000} \\
\hline
\end{tabular}
\end{center}
\label{tbl:eagle}
\end{table}

\begin{table} 
  \caption{The cosmological parameters used from Table 9 in \citet{planck13}:  $\Omega_{\rm m}$,
  $\Omega_\Lambda$, and $\Omega_{\rm b}$ are the average densities of matter,
  dark energy and baryonic matter in units of the critical density at redshift
  zero, $H_0$ is the Hubble parameter, $\sigma_8$ is the square root of the
  linear variance of the matter distribution when smoothed with a top-hat filter
  of radius $8~h^{-1}{\rm Mpc}$.
  }
\begin{center}
\begin{tabular}{|l|r|}
\hline
   $\Omega_{\rm m}$             &    0.307\phantom{00} \\
   $\Omega_\Lambda$             &    0.693\phantom{00} \\
   $\Omega_{\rm b}$             &    0.04825\phantom{} \\ 
   $h \equiv H_0$/(100 km\,s$^{-1}$\,Mpc$^{-1})$ &  0.6777\phantom{0} \\
   $\sigma_8$                  &    0.8288\phantom{0} \\
\hline
\end{tabular}
\end{center}
\label{tbl:cosmo}
\end{table}

Below we summarise the characteristics of \eg, \gl~and \lgl. At the end of this section we have also included a description of the limits in stellar mass used in this paper for both the whole galaxy population and those separated into star-forming and passive galaxies. It is important to
note again that the most fundamental difference between hydrodynamical simulations
and SA models is that the former tracks simultaneously the evolution of dark matter, gas and stellar particles, while
the latter follows the evolution of baryons in haloes in an idealised way.  


\subsection{\eg} 
 \label{sec:eg}

The \eg~simulation suite comprises a set of runs with different box sizes and mass
resolutions. Many of its derived properties are now publicly available \citep{eagleDB}. Here we show results from the largest \eg~simulation
\citep[introduced by][]{schaye15}. The simulation was performed with a version of the \textsc{gadget} code
\citep[last described by][]{Springel2005}, modified by using a state-of-the-art formulation of smoothed particle hydrodynamics (SPH), {\sc anarchy} \citep[Dalla Vecchia {\it in prep.},][]{schaller15}, and subgrid models for galaxy formation.

Halos in the \eg~simulation were identified using
the Friends-of-Friends (FoF) algorithm \citep{Davis1985} and self-bound structures
within the haloes were then identified using the \subf~code
\citep{Springel2001, Dolag2009}. In \eg, the galaxies are defined as the baryonic component of the gravitationally bound subhaloes. Below, we briefly describe the main aspects of the
subgrid physics and some of the definitions relevant here.

\subsubsection{Subgrid physics}\label{sec:eg:sfr}
\medskip\noindent{\bf Star formation}

\noindent Star formation is implemented following the method of \cite{Schaye2008}.
The star formation rate per unit mass
of particles is computed using an analytical prescription designed to reproduce the observed
 \cite{Kennicutt1998} relation in disc galaxies \citep{Schaye2008}. 
Gas particles are converted into star particles stochastically. The threshold in
hydrogen density required to form stars is metallicity dependent, with
lower metallicity gas having a higher threshold. Thus, trying to capture the
metallicity dependence of the phase transition from warm atomic to cold molecular gas \citep{Schaye2004}.

\medskip\noindent{\bf Metal enrichment and gas cooling}

\noindent The simulations assume a \cite{cah03} stellar IMF
 in the range $0.1~\msun$ to $100~\msun$, with each individual star
particle representing a single stellar population. 
The star particles release metals into the ISM gradually over their lifetime through three evolutionary channels: type Ia supernovae, winds and supernovae from massive stars, and AGB stars using the method introduced by \cite{Wiersma2009b}. The yields for each  process are taken from
\cite{Portinari1998}, \cite{Marigo2001} and \cite{Thielemann2003}. The eleven elements that dominate the cooling curve are tracked individually as proposed by \cite{Wiersma2009a}. Assuming that the gas is in ionisation equilibrium, these elements are used to compute the cooling and photo-heating rates of the gas in the presence of the Cosmic Microwave Background and the X-ray and UV backgrounds from galaxies and quasars given by the model of \cite{Haardt2001}.

\medskip\noindent{\bf Feedback from star formation and AGNs}

\noindent Over the course of its lifetime, a simple stellar population will inject energy into the ISM. In \eg, this energy is multiplied by an efficiency factor that depends on the local gas metallicity and density \citep{crain15}. The energy from the stars is transferred to the surrounding gas in the form of heat. The temperature of the surrounding gas is raised instantly by
$10^{7.5}~\rm{K}$.  This is implemented stochastically on one
or more gas particles in the neighbourhood of the star
\citep{DallaVecchia2012}. This gas, once heated, remains coupled in a
hydrodynamic sense with its SPH neighbours in the ISM, and therefore
exerts a form of feedback locally that can directly affect radiative cooling and
star formation. Galactic winds develop without imposing a pre-defined
mass-loading or direction and without disabling radiative cooling.

Supermassive black hole seeds with mass $10^{5}h^{-1}\msun$ are injected in halos above $10^{10}h^{-1}\msun$ \citep{Springel2005} and grow through mergers and accretion of low angular momentum gas
\citep{RosasGuevara2013,schaye15}. AGN feedback
depends on the mass accreted by the black hole and is modelled by the injection of thermal
energy into the gas surrounding the black hole such that its temperature is raised by $\Delta T=10^{8.5}~K$
\citep{Booth2009,DallaVecchia2012}.

\subsubsection{Aperture measurements}
The stellar mass of a galaxy in the \eg~simulation is defined to be the
    sum of the masses of all the stellar particles that are bound to the
    corresponding subhalo within a 3D aperture of radius 30 proper kiloparsec
    \citep{schaye15}. The stellar mass computed in this way is found to be similar to the mass computed within Petrosian apertures from the simulation at $z=0.1$. Meanwhile, in SA models, the stellar
    mass is accumulated with time, starting from the initial seed of hot gas in a halo and taking into account the fraction of mass returned to the ISM by SNe and
stellar winds for a simple stellar population and in the case of \lgl~the losses due to tidal disruption.

For consistency with the galaxy mass definition, star formation rates (SFRs) of galaxies in \eg~are measured within spherical apertures of 30~proper kpc. As the majority of the star formation occurs close to the centres of galaxies, this aperture has a negligible effect on the total SFR recovered.

\subsubsection{Calibration of the parameters}

As discussed by \citet{schaye15}, the free parameters controlling
the subgrid model for feedback were chosen in order to reproduce the stellar mass
functions at $z\approx 0$ from the GAMA survey by \citet{baldry12}
 and from the SDSS survey by \citet{liw09}, the galaxy mass-size relation as reported by \citet{shen03} and \citet{baldry12} and the relation between the mass of the central supermassive black hole and the total stellar mass of galaxies
derived from observations compiled by \citet{mcconnell13} (see also Table~\ref{tbl:cali}). 

\subsection{Semi-analytical models}\label{sec:sams}

Semi-analytical (SA) models use simplified, partly phenomenological recipes and
rules to follow the fate of baryons in a dark matter dominated universe in which structure grows
hierarchically through gravitational instability \citep[see][for an overview
of SA models]{bau06,benson10}. 

\gl~\citep{cole00,bow06,gp14,lacey15} and \lgl~\citep{Springel2005,croton06,qi11,henriques14}, the two models used for this
study, follow the physical processes that shape the formation and evolution of
galaxies, including: 
\begin{enumerate}
  \item the collapse and merging of dark matter haloes; 

  \item the shock-heating and radiative cooling of gas inside dark matter
    haloes, leading to the formation of galaxy discs;

  \item star formation bursts that can be triggered either by mergers or disc
    instabilities;

  \item the growth of supermassive black holes in galaxies;

  \item feedback from SNe, from AGN and from photoionisation of the intergalactic medium; 

  \item chemical enrichment of stars and gas, assuming the instantaneous
    recycling approximation (as opposed to \eg, where a non-instantaneous recycling is implemented);

  \item galaxy mergers driven by dynamical friction within common dark matter
    haloes, leading to the formation of stellar spheroids, which can also be produced by disc instabilities.

\end{enumerate}
The models also compute the sizes of the disc and bulge components of galaxies.
The end product of the calculation is a prediction for the abundances and
properties of galaxies that reside within dark matter haloes of different
characteristics.

In order to make a fair comparison
with \eg, the two SA models which we use here have been adapted from the published
models described by \citet{qi13} and \citet{gp14}. Specifically, the SA models have
been run on merger trees extracted from the \egdm~simulation, assuming the same
Planck cosmology (Planck Collaboration 2014, Table 9) as adopted by \eg. Both
the underlying simulation and the cosmology are different from the WMAP7
cosmology used by \citet{qi13} and \citet{gp14}, and thus, a recalibration of
their free parameters was required in order to satisfactorily reproduce the
corresponding set of observational data summarised in \S\ref{sec:calibration}.
Moreover, \eg~assumes a Chabrier IMF \citep{cah03}, which is the default in
\citet{qi13} but not in the published \citet{gp14} model. Thus, there is a
corresponding change in the metal yield and recycled fractions in the SA
models (see \ref{app_cal} for further details). For both \eg~and \lgl, galaxy photometry has been derived using the
stellar population synthesis models of \citet{Bruzual03}. In the case of \gl,
the models of \citet{cw10} are used, which include a very similar library of
stellar spectra to \citet{Bruzual03} and also account for the contribution of
thermally pulsating asymptotic giant branch stars \citep[see][for a comparison
of different stellar population synthesis models]{gp14,Trayford2015}. 

The most significant difference between the published \gl~model and that presented here by default, is the inclusion of gradual ram pressure stripping of hot gas in the satellite galaxies of the \gl~model, as opposed to the instantaneous stripping assumed by \citet{gp14} (see \S\ref{sec:samp}). This was included to ensure all three models provided a reasonable match to the observed passive fractions at $z=0$. Note that the \citet{gp14} model has previously been used including such an update of the hot gas stripping, in the context of studying early-type galaxies \citep{lagos14}. Throughout this study, we comment on the extent of the effect of stripping the hot gas instantaneously or gradually, for results that are significantly affected by this choice.

An overview of the \gl~and \lgl~models is provided below, focused on the aspects where they differ.

\subsubsection{Halo finder and merger trees}
  \label{sec:mt}

Both SA models are based on merger trees extracted from the same \egdm~simulation,
however, there are differences in the methods applied to construct them. \gl~is
based on subhalo merger trees built with the Dhalo algorithm \citep{jiang14}, while
\lgl~subhalo merger trees are constructed following \citet{Springel2005}, \citet{dlb07} and \citet{boy09}. As in \eg, both methods use the FoF algorithm to identify haloes, but ensuring that haloes artificially linked by this algorithm are treated as separate objects. \subf~is used to identify the self-bound substructure in haloes. Initial \subf~merger trees are built by tracking particles between snapshots. Both methods can identify the descendants of a
halo at any of the following 2 snapshots in the case of \lgl~and 5 in the Dhalo
algorithm. This feature was implemented in order to improve the identification
of substructure in close encounters that can be mistaken for real mergers. In effect, the only conceptual difference between the two methods is that the Dhalo algorithm enforces a monotonic growth of halo mass.

In Appendix~\ref{appendix:halomf}, we show the halo mass functions from the
three models considered, focusing on variations that arise from the different definitions of halo mass.
Although systematic discrepancies do exist due to these definitions, the dominant difference is between the \eg~and
the \egdm~simulations, due to the impact that early
expulsion of baryons has on the subsequent growth of dark matter haloes
\citep[e.g.][]{sawala13,velliscig14,schaller14}. 

\medskip\noindent {\bf Central and satellite galaxies}

\noindent For \gl, host haloes are defined either at the final output of the simulation or just before a halo merges with another more massive one. The centre of the most massive subhalo is defined as the halo centre. The Dhalo algorithm determines the main progenitor of this subhalo as the one that contributed the most bound part of the descendant. This process is carried out
starting at the final output time and working backwards towards high redshift.
This results in the assignment of one central subhalo to every halo in such a
way that the same subhalo is considered to be the central as long as the halo
exists \citep[see appendix A of][]{jiang14}.

In \lgl, as in \eg, central galaxies are those hosted by the most massive subhalo (main
subhalo) which usually has most of the mass of its FoF group. This choice for defining the main branch, tries to reduce the chance of the centre swapping to a different subhalo between snapshots \citep{eagleDB}. 

In \gl, satellite galaxies remain as such until either they merge with the central galaxy in their host halo or the end of the simulation is reached. This is not the case in \lgl, in which satellite galaxies can be reclassified as centrals if they are far enough from the virial radius of the halo that was hosting them.

\subsubsection{Physics}\label{sec:samp}

\noindent{\bf Star formation}

\noindent In \gl, the cold gas corresponds to the ISM gas, including the molecular, atomic
and ionised phases. In this model, the quiescent star formation in galaxy discs
explicitly depends on the molecular component of the gas \citep{lagos11}.  This
empirically motivated calculation assumes that during quiescent star formation,
the surface density of the SFR is proportional to the surface density of
molecular hydrogen in the ISM \citep{blitz06,leroy08,bigiel08}. The SFR
from star-bursts in \gl~is assumed to be simply proportional to the total mass of
cold gas present in galaxy bulges and inversely proportional to a star formation
time-scale \citep{granato00}. 

In \lgl, stars are assumed to form from the gas in the quiescent mode. The cold gas
disc and stellar disc in the model are distinct, and both can grow continuously in mass and angular momentum in a physically-plausible way
\citep[][\S 3.3]{qi13}. Stars form in a cold gas disc  according to a simplified
empirical Kennicutt relation \citep{ken98}, but only in regions where the
surface gas density exceeds a critical value. This critical value is related to the gas
velocity dispersion and the rotation
curve of the galaxy. This star-formation threshold reflects that the star formation is expected to be possible only in dense enough
regions \citep{kauffmann96}. The SFR from star-bursts in \lgl~is assumed  to be proportional to the total mass of cold gas and the mass ratio of two merger progenitors whenever merger happens. 


\medskip\noindent{\bf Stellar feedback}

\noindent When massive stars die, they inject large amounts of energy into the ISM in SN explosions.  In both \gl~and \lgl, this can cause ejection of gas from galaxies and halos, but the details are different. 
In both models, cold gas is ejected from galaxies at a rate proportional to the SFR, with the proportionality factor (called the mass-loading factor) depending on the circular velocity. 
In \gl\ this dependence is a simple power law, while in \lgl\ it has a more complicated form, but in both models, the mass-loading factor decreases with increasing circular velocity. Furthermore, in \gl\ the circular velocity used is that at the disc half-mass radius for disc star formation, and at the bulge half-mass radius for starbursts, while in \lgl\ it is the peak circular velocity of the subhalo. 
In \gl, cold gas is ejected directly from the galaxy out of the halo. In \lgl, there is instead a two-stage ejection process: cold gas is ejected into the halo, and SN also inject energy into the halo, with an efficiency that also depends on circular velocity; hot gas is then ejected from the halo in a quantity depending on the available energy, with an explicit constraint that the energy used cannot exceed the total SN energy.

In both models, gas that has been expelled from the halo is added to a reservoir outside the halo, from where it gradually returns to the hot halo, being reincorporated at the halo virial temperature. In \gl\ the return timescale is simply proportional to the halo dynamical time, while in \lgl\ it also depends on the halo circular velocity, with the return timescale decreasing with increasing circular velocity.



\medskip\noindent{\bf AGN feedback }

\noindent The onset, by AGN activity, of the suppression of the gas cooling in haloes is assumed to occur in both \lgl~and \gl~only for haloes hosting galaxies whose central black hole is growing in mass through gas
accretion.

In \gl, the quasi-hydrostatic cooling is assumed to occur in haloes hosting
galaxies such that $t_{\rm cool}>t_{\rm ff}/\alpha_{\rm cool}$, where $t_{\rm
cool}$ is the cooling time of the gas, $t_{\rm ff}$ is the free-fall time for the
gas to reach the centre of the halo and $\alpha_{\rm cool}$ is a model
parameter, set to $\alpha_{\rm cool}=0.52$. $\alpha_{\rm cool}$ is set to 0.6 in both the
published version \citep{gp14} and the version with an instantaneous stripping
of hot gas in satellite galaxies shown in this study. When a halo is undergoing quasi-hydrostatic cooling, the gas cooling
is suppressed if the luminosity released by gas accreted onto a central
supermassive black hole balances or exceeds the cooling luminosity
\citep[see][for further details]{bow06,fanidakis11}. 

In \lgl, it is assumed that the hot-mode accretion of mass onto the BH 
deposits energy with a 10\% efficiency, heating up the halo hot gas. The black hole accretion rate in this model is assumed to be a function of the ratio of hot gas mass to subhalo DM mass, the
virial velocity of the halo and the mass of the central black hole. The
efficiency of the growth of BHs due to such hot-mode accretion is a model
parameter. BHs can also grow by mergers \citep[see][for further details]{croton06,qi11}.

\medskip\noindent{\bf Sizes} 

\noindent For the calculation of disc sizes, both SA models assume conservation of specific angular momentum and centrifugal equilibrium. The sizes of spheroids are estimated in both models assuming virial equilibrium and energy conservation.
To determine disc sizes, \lgl~follows the full angular
momentum vectors of haloes and discs, and separates the contribution from stars and gas in the disc \cite[][\S 3.3]{qi11}, while \gl~only tracks the magnitudes of the disc and halo angular momentum, assuming that the disc and halo angular momentum are always aligned \citep[][\S 4.4 \& Appendix C]{cole00}. In both models the disc angular momentum is determined
by the halo formation and gas cooling history. In \gl, this is then used to obtain both the disc radius and
the circular velocity at the disc half mass radius by solving self-consistently the combined gravity of the disc, spheroid and
halo \citep{cole00}. In \lgl~the
circular velocity of the disc is assumed to be equal to the maximum circular velocity of the host halo \citep{qi11}. 

\gl~includes the self-gravity of discs and spheroids when computing
disc sizes, while \lgl~ignores this, which is a significant assumption, in particular for massive galaxies whose inner
regions are, in principle, not dominated by dark matter. Furthermore,
\gl~also models the contraction of the dark matter halo due to the
gravity of the baryonic component. We note that if the baryonic self-gravity was turned off in \gl~then the circular velocity used by this model would be close to the maximum halo circular velocity, as assumed by \lgl, because in that case, \gl~would use the uncontracted dark matter halo value of the circular velocity at the half mass radius of the disk.

Although the gravity of the baryons
should be taken into account when calculating the distribution of dark
matter in a halo, the simplified model for halo contraction adopted by
\gl~appears to overestimate the effect of the baryons compared to gas
dynamical simulations \citep[e.g.][]{gnedin04}, this maybe because the adiabatic invariance assumed in the contraction model is violated by the short timescale of SN driven outflows in low mass haloes \citep{sawala13,newman15}.

\medskip\noindent{\bf Environmental processes}

\noindent In both \lgl~and \gl, environmental effects, such as ram-pressure stripping
of gas, are implemented in a way that only impacts the evolution of satellite galaxies. Note that these environmental effects are naturally included in hydrodynamical simulations such as \eg. The SA models used in this work assume a gradual ram-pressure stripping of
the hot gas in satellite galaxies. \lgl~also includes a
basic model of tidal stripping. In this model, the hot gas in a subhalo is distributed following the underlying dark matter and it is affected in the same way as the dark matter by tidal stripping. Once a subhalo has been entirely disrupted, the remaining galaxy will be disrupted when the baryon density within its half-mass radius is smaller than the main halo density at the pericentre of the subhalo orbit. The components of the disrupted satellite galaxy are then assigned to a
population of intracluster stars \citep{qi11}. 

Many of the \gl~published models, including \citet{gp14}, adopt instantaneous ram-pressure 
stripping, as opposed to gradual stripping, of the hot gas in satellite galaxies \citep[but
see][]{lagos14,lagos15grp}. Here we use the parametrisation for the
gradual stripping of hot gas in satellites introduced by \citet{font08} following the analysis of the
hydrodynamical simulations of cluster environments by \citet{mccarthy08}.
Assuming instantaneous stripping results in the exhaustion of most of the satellite galaxy gas
reservoirs, quickly quenching their star formation, as no further supply of gas
is accreted. Thus, most satellites in models with instantaneous stripping are passively evolving. The assumptions made about the gas in
satellite galaxies affect the results related to separating galaxies into
star-forming and passive subsets. However, the change in the ram pressure stripping model has only a small effect on other results, such as the calibration diagnostics (see \S\ref{sec:calibration}). 

Although there is plenty
of observational evidence indicating the importance of gas stripping for galaxies
within dense environments \citep[e.g.][]{scott13,fumagalli14,boselli14}, the modelling of this process is unclear. One of the primary uncertainties is related to the fate of the stripped gas once it has been ejected from the subhalo by stellar feedback. Another concern is that, as observations are limited to cluster environments and to ram-pressure
stripping of the ISM (and not the ram-pressure stripping of hot gas), galaxies in lower density environments in SA models might be over-quenching the star formation \citep{hirschmann14,mcgee14}. We will investigate the role of ram pressure stripping through the comparison with \eg~and between the \gl~models with instantaneous and gradual stripping. In the subsequent discussion we comment on the
instantaneous stripping \gl~model when significant differences arise relative to the default \gl~model.

\subsubsection{Calibration}\label{sec:calibration}
\begin{table*} 
  \caption{The observational data used to calibrate the three default models used in
  this work.}
\begin{center}
\begin{tabular}{|l|l|}
\hline
    Model  & Observational data used in the model calibration\\
  \hline
  \eg~ & Galaxy stellar mass function at $z\approx 0.1$ from GAMA \citep{baldry12} and SDSS \citep{liw09},\\
       & the stellar mass-size relation at $z\approx 0.1$ \citep{shen03,baldry12} \\
       & and the $z\approx 0$ $M_{\rm BH}-M_*$ relation \citep{mcconnell13}. \\
   \gl~&  $b_{J}$-band {\citep{norberg02}} and $K$-band {\citep{driver12}} 
                      luminosity functions at $z\approx 0$, \\
                      &the passive fraction at $z\approx 0$ \citep{gilbank2011,bauer2013} and the \\
                      & $M_{\rm BH}-M_{\rm Bulge}$ relation \citep{haering.rix04}. \\
   \lgl~ & Galaxy stellar mass function at $z\approx 0$ \citep{baldry08,liw09} and the \\
                      & $M_{\rm BH}-M_{\rm Bulge}$ relation \citep{haering.rix04}. \\
\hline
\end{tabular}
\end{center}
\label{tbl:cali}
\end{table*}

For \gl, the \citet{gp14} model was calibrated to reproduce the $b_{J}$ and $K$-band luminosity functions at $z=0$ from \citet{norberg02} and \citet{driver12} and the black hole- bulge stellar mass ($M_{\rm BH}-M_{\rm Bulge}$) relation from \citet{haering.rix04}, see also Table~\ref{tbl:cali}. As mentioned earlier, here we are using a modified version of this model, for
which we have adopted the Planck cosmology, the \citet{cw10} stellar population synthesis model and a Chabrier IMF, as opposed to the
previously used \citeauthor{kennicutt_imf} IMF, with the corresponding values
for the yield and recycled fraction. These modifications did not significantly
alter the $z=0$ luminosity functions used for calibrating the free parameters in
the model. As mentioned, the instantaneous stripping of hot gas in satellite galaxies is replaced with a gradual stripping model. The predicted passive fraction for the model with gradual stripping is closer to that observed \citep{gilbank2011,bauer2013}. The change in the $b_{J}$ and $K$-band luminosity functions at
$z=0$ resulting from assuming gradual stripping is large enough to require a slight lowering of the threshold for the AGN feedback to be effective and recover the same level of agreement with the observed luminosity functions. The individual impact of each of these variations are discussed in detail in Gonzalez-Perez et al., in prep.

The published \citet{qi13} model was calibrated primarily to reproduce the $z\approx 0$
stellar mass function observed by \citet{baldry08} and
\citet{liw09} and the $M_{\rm BH}-M_{\rm Bulge}$ relation of \citet{haering.rix04}, as is summarised in Table~\ref{tbl:cali}. During the calibration of this model a further condition ensured that the cold gas fractions increase with decreasing stellar mass, as observations suggest. The
\lgl~model used in this work has been recalibrated such that it still reproduces the aforementioned observations by slightly modifying the stellar and AGN feedback in order to account for the change in cosmology, halo mass resolution and time sampling of the merger trees, that arise as a result of the model being built on the \egdm~simulation. 

\subsection{Stellar mass limits and star-forming galaxies}\label{sec:sf_def}
In order to reduce the sampling effects associated with the limited resolution of the \eg~simulation, only galaxies with a minimum stellar mass $M_*\gtrsim 10^{8}~\msun$ are considered. We impose this cut in stellar mass in the three models. Moreover, in the \eg~simulations, galaxies with low star formation rates can present quantised behaviour in the sense that a SN explosion in a single stellar particle can modify the star formation by a significant amount, due to poor sampling. Thus, a minimum number of about 30 star-forming particles is needed in order for the star formation rate (SFR) to be reliable, based on resolution tests from \citet{schaye15} at low and \citet{fur14} at high redshifts. This limit is shown by the sloping cyan lines in Fig.~\ref{fig:ssfr_m}. 

In this work we separate passive from star-forming galaxies based on their
specific star formation rate (sSFR$=$SFR$/M_*$). The chosen boundaries are
highlighted in Fig.~\ref{fig:ssfr_m} (horizontal dashed cyan): $\log_{10}({\rm
sSFR}/{\rm Gyr}^{-1}) = -2, -1.04, -0.97$ at redshifts $z=0,\, 1,\, 2$
respectively. Galaxies above these cuts are considered to be star-forming and
galaxies below are considered as passive. \citet{fur14} set these limits, which
correspond to $\approx 1$~dex below the mean sSFR from a compilation of observed
star-forming galaxies. We have tried different values of the sSFR cut used to
split galaxies into star forming and passive populations, including those from
\cite{Franx08}. Although some of the results are quantitatively affected by the
exact value of this cut, such as the passive fractions, the discussion and
conclusions in this paper are insensitive the particular value choosen, within
one sigma of the values stated above.


Note that the sSFR value chosen as a boundary for
separating galaxies into passive and star-forming intersects with that
corresponding to the minimum of 30 star-forming particles at a stellar mass that
decreases with increasing redshift. Thus, the minimum stellar mass for measuring
SFRs in \eg~varies with redshift as an indirect consequence of imposing a
boundary between passive and star-forming galaxies that evolves with redshift.

\section{Stellar masses}\label{sec:sm}

Many aspects of galaxy evolution are condensed
into the galaxy stellar mass function (GSMF) and related quantities. In this section,
we compare different stellar mass relations obtained with \eg, \gl~and \lgl,
for a selection of redshifts. 

\subsection{The galaxy stellar mass function}\label{sec:smf}
\begin{figure}
  \includegraphics[width=3.0in]{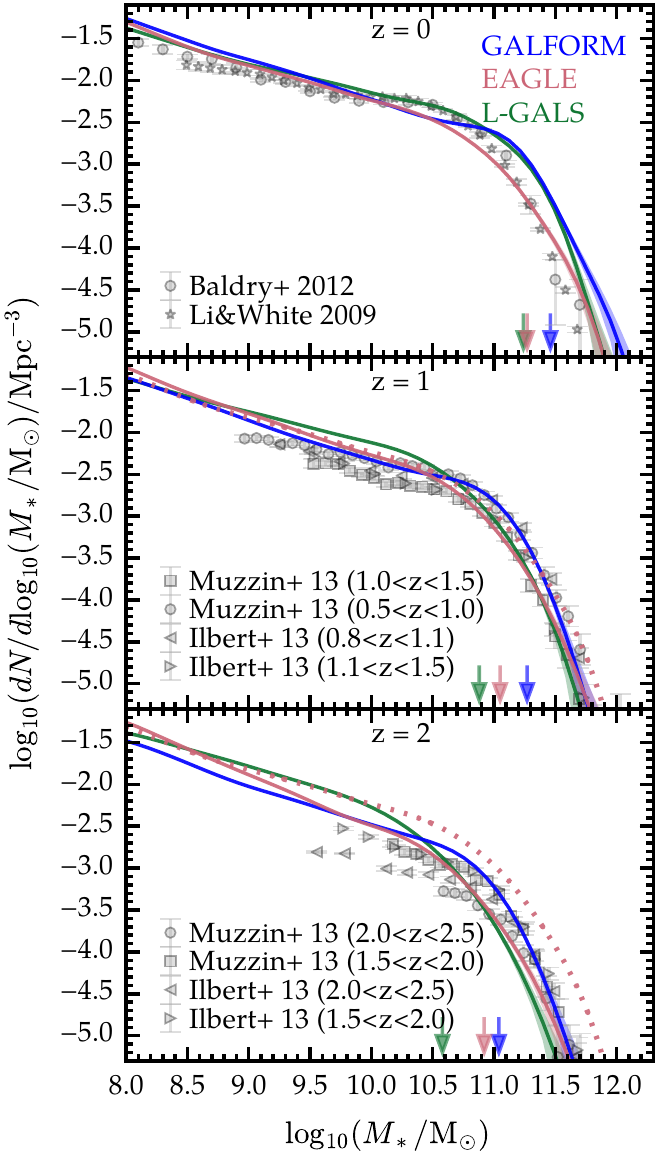}
  \caption{\label{fig:smf}  Galaxy stellar mass functions in \eg, \gl~and
  \lgl~at different redshifts, as indicated by the legend, convolved with a Gaussian error of $0.07+0.04z$ \citep{Behroozi13}. The shaded regions
  show the 1$\sigma$ error from 200 bootstrap samples of galaxies taken from the whole
  simulation volume (note that this is a very narrow region for most stellar masses). The arrows show, in the same colour as the corresponding
  model, $\mb$ from a single Schechter function fit to the GSMFs in the range
  $10^{8}<M_*/\msun<10^{12.5}$, using the 1$\sigma$ errors shown. The \eg~stellar mass function at $z=0$ is overplotted as a red dotted line in
  the middle and lower panels. For reference, observational estimates of the GSMF from \citet{liw09,baldry12,Muzzin2013,Ilbert2013} are included as indicated in the legend (note that when necessary, the observational data have been converted to a \citeauthor{cah03} IMF and the Planck Cosmology). The evolution of the GSMF is remarkably similar
  for the three models.
} 
\end{figure}

   
\begin{table}
  \caption{The model GSMFs for \eg, \gl~and \lgl, at three different redshifts, have been fitted to a single Schechter function, $\Phi(M) = \Phi^{*}e^{-(M_* - \mb)}(M_* - \mb)^{1+\alpha}$, in the range $10^{8}<M_*/\msun<10^{12.5}$ and using the 1 $\sigma$ errors shown in Fig.~\ref{fig:smf}. For these fits, the table below presents the faint-end slope, $\alpha$, the stellar mass at the knee of the GSMF, $\mb$, and two measures of the goodness of the fit: $\chi^2$ and that normalised by the degrees of freedom, $\chi^2/\nu$.}
\begin{center}
\begin{tabular}{|l|c|c|c|c|}
\hline
\eg~& $\, \alpha$ & $ \log_{10}(\mb/\msun)$ & $ \chi^2$ & $\chi^2/\nu$ \\
    \hline
    $z=0$  & $-1.43$ &  $11.2$ & 0.047 & 0.0011      \\
    $z=1$  & $-1.49$ &  $11.1$ & 0.079 & 0.0020     \\
    $z=2$  & $-1.60$ &  $10.9$ & 0.064 & 0.0018     \\
\hline
\gl~& & &  \\
    \hline
    $z=0$  & $-1.44$ &  $11.4$ & 0.396 & 0.0084      \\
    $z=1$  & $-1.45$ &  $11.3$ & 0.718 & 0.0167     \\
    $z=2$  & $-1.47$ &  $11.0$ & 0.421 & 0.0105     \\
\hline

\lgl~& & & \\
    \hline
    $z=0$  & $-1.34$ &  $11.2$ & 0.144 & 0.0036      \\
    $z=1$  & $-1.35$ &  $10.9$ & 0.012 & 0.0003     \\
    $z=2$  & $-1.40$ &  $10.6$ & 0.084 & 0.0025     \\
\hline
\end{tabular}
\end{center}
\label{tbl:fitting}
\end{table}

In Fig.~\ref{fig:smf} we show the GSMF\footnote{Throughout this
paper, we present model distributions estimated by the standard Kernel Density
Estimation with bandwidth of 0.2 \citep{silverman86} rather than histograms. This choice minimises the dependence on the chosen starting point that simple histograms have.} of model galaxies in \eg, \gl~and \lgl~at three different redshifts,
$z=0,\, 1,\, 2$. The GSMF can generally be described approximately by a Schechter function\footnote{Other functional forms might be more appropriate
than the Schechter function for describing either the mass or luminosity
functions \citep[e.g.][]{gun15}, in particular for cases such as the GSMF
predicted by the \gl~model, which presents a plateau just below $\mb$.} \citep{schechter76}, i.e. a
power law and an exponential break which starts at a characteristic mass, $\mb$. We carry out single Schechter function fits, using the Levenberg-Marquardt algorithm through least squares, to each model at all redshifts in the stellar mass range $10^{8}<M_*/\msun<10^{12.5}$. The best-fitting parameters are given in Table~\ref{tbl:fitting}. This provides a quantitative summary of the results shown in Fig.~\ref{fig:smf}. 

The GSMFs at $z=0$ are similar for the three models. The parameters in \eg~and \lgl~were calibrated to reproduced the observed GMSF at $z=0$, and \gl~to reproduce the observed K-band luminosity function at $z=0$, which follows closely the GMSF. Thus, by construction, at $z=0$, the $\mb$ obtained from a single Schechter function fit to the
GSMFs from the three models is similar, with a
variation $\lesssim5$\% (see Table~\ref{tbl:fitting}). For galaxies with
$M<10^{10.5}\msun$, the number densities in the three models are similar, with
differences $\lesssim0.3$~dex. However, in \eg~the GSMF at the fitted $\mb
\approx 10^{11.2}~\msun$ is a factor of $\approx 3$
below those from the SA models. From exploring the parameter space with the SA models, it is
clear that both the normalisation and the position of the knee in the GSMF are
mostly affected by stellar and AGN feedback \citep[e.g.][]{bower12}. Note that \eg~was calibrated using small volumes, which do not constrain galaxies with $M_* \gtrsim \mb$ however the results are still consistent with observations within the (systematic) errors \citep{schaye15}.

The three models predict the GSMF to 
evolve in similar ways, such that both the overall GSMF normalization and
$\mb$ decrease with increasing redshift. The single
Schechter function fit to the GSMFs in the range $10^{8}<M_*/\msun<10^{12.5}$, with the 1 $\sigma$ errors shown in Fig.\ref{fig:smf}, gives $\log_{10}(\mb/\msun)$ which varies in
a similar way for the three models, as indicated by the arrows in Fig.\ref{fig:smf} (see also Table~\ref{tbl:fitting}). For all models, the decrease in abundance is most significant for
massive galaxies. The faint-end slope becomes slightly steeper with redshift in all three models (see Table~\ref{tbl:fitting}) and this evolution is strongest for \eg. We have extended the comparison of model GSMFs up to $z=4$ (not shown), finding similar trends to those reported here up to $z=2$.

As mentioned in \S\ref{sec:samp}, the \gl~model shown in Fig.~\ref{fig:smf}
assumes that the hot gas in satellites is removed gradually as opposed to
instantaneously, as assumed in the previously published model of
\citet{gp14}. As we show in the next section, around $\mb$ the GSMF is
dominated by central galaxies. These grow faster when an instantaneous
ram-pressure stripping of the hot gas is assumed, at least partly because the
central galaxy halo gains more gas from the satellite haloes. This has an impact
on the massive end of the GSMF at $z=0$, which is compensated for during calibration by the small change
in the AGN feedback discussed in \S\ref{sec:calibration}.

\subsubsection{The passive and star-forming GSMFs}\label{sec:sf_smf}
\begin{figure*}
  
  \includegraphics[height=2.9in]{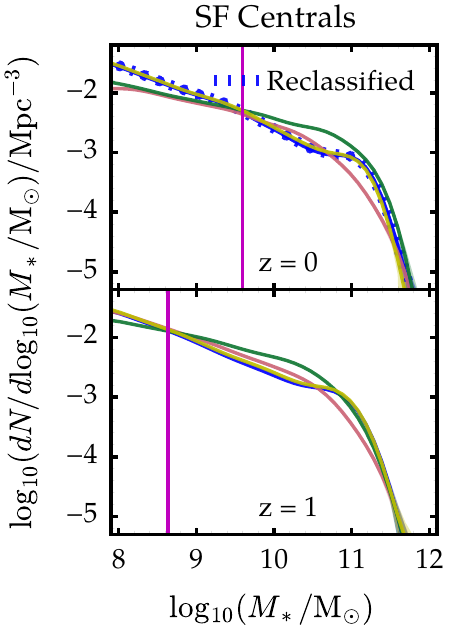}
  \includegraphics[height=2.9in]{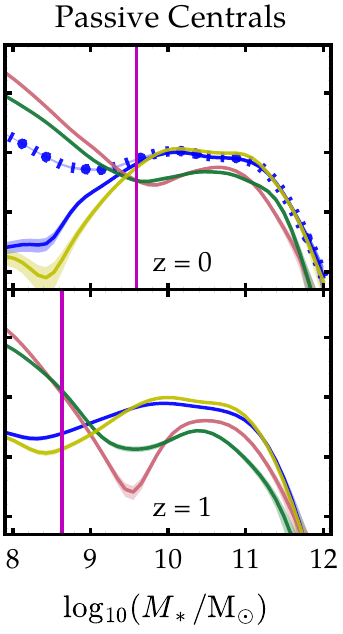}
  \includegraphics[height=2.9in]{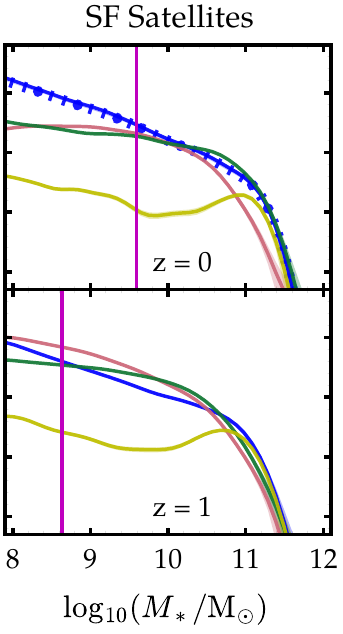}
  \includegraphics[height=2.9in]{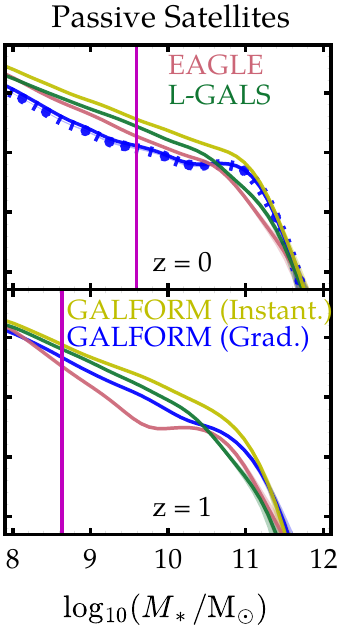}
  \caption{\label{fig:fig5_smf_split} The GSMF at $z=0$, top panels, and
  $z=1$, bottom panels, calculated from \eg, \gl~and \lgl. From left to right the columns show the GSMF of:
  star-forming central galaxies, passive centrals, star-forming satellite
  galaxies and passive satellites. The separation into star-forming and passive galaxies is
  described in \S\ref{sec:sf_def}. The vertical cyan lines are a guide to the
  resolution limits of star-forming galaxies in \eg. The limit evolves with redshift due to the evolving sSFR limit coupled with the need to have at least 30 star-forming particles in \eg~to measure SFRs adequately (see \S\ref{sec:sf_def} for further details). In this plot we show the predictions from \gl~assuming either gradual ("Grad.",
  black lines, the default model in this paper) or instantaneous ram pressure
  stripping ("Instant.", yellow lines) of the hot gas in satellite galaxies. We
  also show, as dotted lines, the result from reclassification as centrals of those satellite galaxies in \gl~that are beyond one virial
radius from the central galaxy of their host halo. This exercise shows that the upturn at low-masses in the passive centrals GSMF for \eg~and \lgl, shown in the second panel from the left, is at least partly due to ejected satellite galaxies, as discussed in \S\ref{sec:sf_smf}.
}
\end{figure*}

In Fig.~\ref{fig:fig5_smf_split} we present the GSMFs for central and satellite
galaxies, separated into star-forming and passive as described in \S\ref{sec:sf_def}. Star-forming galaxies dominate the number density in the global GSMF 
at masses $M_*\lesssim\mb$, while, in the same stellar mass range, centrals dominate over satellite galaxies.

The most striking difference for central galaxies seen in
Fig.~\ref{fig:fig5_smf_split} is the upturn in the abundance at low-masses of passive
centrals predicted by both \eg~and \lgl, but not seen for galaxies in
the \gl~model. At $z=0$, this upturn happens at $\approx 10^{9.5}~\msun$, very
close to the resolution limit of \eg~for star-forming galaxies (see
\S\ref{sec:sf_def} for details). At higher redshifts, the upturn seen for \eg~galaxies is also likely to be due to sampling, given that
\citet{fur14} found the fraction of low-mass passive galaxies to vary with higher simulation resolution. In contrast, \lgl~and \gl~predictions are constrained by the halo resolution of \egdm, which mainly affects galaxies with $M_* < 10^8~\msun$, much lower than the mass at
which the upturn starts for \lgl. In order to explore the possible origin of the
upturn in the abundance of small passive centrals, we ran the \gl~model changing
in turn parameters controlling the gas cooling, star formation law
\citep[including a star formation law with an explicit threshold as described
in][]{lagos11} and feedback. None of these aspects were found to produce an abundance of passive central galaxies close to that from \lgl.


Further investigation revealed that the different behaviour for small masses arises from the definition of central and satellite galaxies in the models (\S\ref{sec:models}). The most relevant
difference here is that only the \gl~model assumes that once a galaxy becomes a
satellite it will remain as such until it merges with a central galaxy. In \gl, satellite galaxies can leave their host halo. When that happens, these galaxies experience ram-pressure stripping, even if outside their host halo. Such long range environmental processes has been observed \citep[e.g.][]{Wetzel2014,hirschmann14}.
This implies that for \lgl, galaxies that are classified as central at $z=0$ could have
undergone environmental quenching processes at earlier times. 
This is actually the case for around a third of the $z=0$ centrals in
the \lgl~model \citep{henriques14,hirschmann14}, which we refer to as
ejected satellite galaxies \citep[][studied the evolution the host
haloes of these type of galaxies]{li13}.  
Hence, these differences between the definition of
satellite galaxies play an important role in the interpretation of results. As a test of the importance of the satellite galaxy definition at $z=0$, we have reclassified satellite galaxies
in \gl~as centrals if they are beyond one virial
radius\footnote{The virial
radius for this test was simply assumed to be related to the mass and circular
velocity of the host halo as:  $R_{\rm vir} =G\times M_{\rm host\, halo}/(V_{\rm
host\, halo}^2)$, with $G$ being the gravitational constant.} from the central galaxy of the host halo. This results in only $\approx 5$\% of the
satellites at $z=0$ being reclassified. At $z=0$, the resulting GSMFs from this reclassification are shown in Fig.~\ref{fig:fig5_smf_split} (dotted lines) and are found to present an upturn at $M_*\approx 10^{9}~\msun$, as expected. The effect of this
reclassification exercise is negligible for galaxies that are not passive centrals.

\gl~predicts a larger number of low-mass star-forming galaxies than the other models at $z=0$. This suggests that either the stellar feedback in \gl~is weaker or the reincorporation times for the gas are shorter than in the other two models, however, some other physical process might be relevant given that at higher redshift this excess is not clearly seen.

In Fig.~\ref{fig:fig5_smf_split} we present two flavours of the \gl~model: the default model described in
\S\ref{sec:sams}, which assumes a gradual stripping of the hot
gas in satellite galaxies, and one assuming the stripping to be instantaneous, as used in many previous \gl~publications \citep[e.g.][]{gp14}. The two right-hand panels in
Fig.~\ref{fig:fig5_smf_split}, which show satellite galaxies, reveal that the SA models assuming gradual
stripping predict GSMFs are closer to those predicted by \eg. Instantaneous stripping results in exhausting most of satellites gas reservoir and a quick quenching of the star formation in these galaxies, as no further supply of gas is accreted onto satellite
galaxies. Thus, most satellites in the \gl~model with instantaneous stripping
are passively evolving, as can be seen in the right panels of
Fig.~\ref{fig:fig5_smf_split}. The assumptions made about the gas in satellite
galaxies affect the results related to the split of galaxies into
star-forming and passive types. 

While in the SA models the ram-pressure stripping is only modelled for the hot
component in satellite galaxies, in \eg, as in other hydrodynamical simulations,
the environment affects both cold and hot gas in all galaxies \citep{bahe13}. A
detailed study of the differences between SA models and hydrodynamical simulations
will require the comparison of individual galaxies in similar (if not the same) haloes,
something that is beyond the scope of this paper.

\subsection{The build-up of the stellar mass}\label{sec:build_smf}

\begin{figure}
  \includegraphics[width=3.0in]{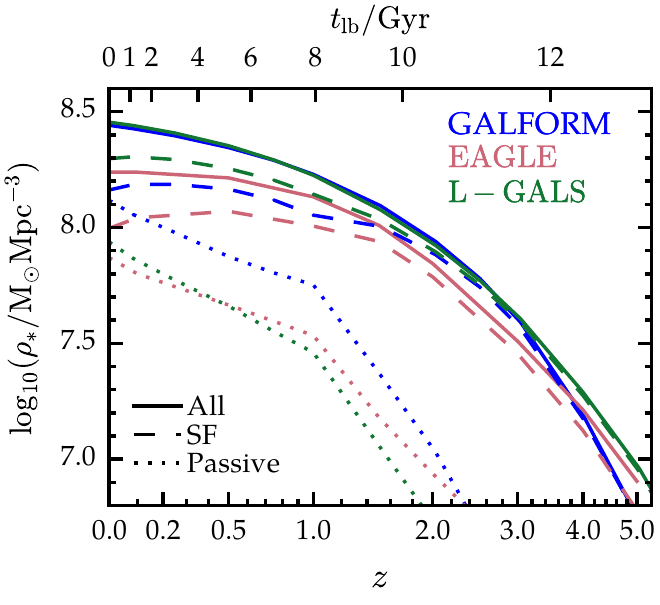}
  \caption{\label{fig:sm_vs_z} The stellar mass density of galaxies with $M_*>10^8~\msun$, as a function of redshift and lookback time, $t_{\rm lb}$ Gyr, as
  predicted by \eg, \gl~and \lgl. Solid lines show the prediction for all
  galaxies (note that the corresponding lines for \gl~and \lgl~are practically on top of each other), dashed lines show star-forming galaxies only and dotted lines passive galaxies only. The star-forming and passive galaxies are classified
  according to the boundary values of the sSFR described in \S\ref{sec:sf_def}, which are
  interpolated linearly in redshift. The global stellar mass density at $z<1.5$ is dominated by galaxies with $M_*\ge
10^{10.5}~\msun$.}
\end{figure}

The build-up of the GSMF is closely related to the evolution of the comoving stellar mass
density, which is presented in Fig.~\ref{fig:sm_vs_z} for model
galaxies with stellar masses above $10^8~\msun$. In turn, given that the two SA models adopt the
instantaneous recycling approximation, their stellar mass density at a given redshift, $\rho _* (z_i)$, can be recovered by integrating the star formation rate comoving density (SFRD),
$\dot{\rho _*}$ (shown in Fig.~\ref{fig:sfrd_vs_sm}), and subtracting stellar mass
losses:
\begin{equation}
\rho _* (z_i) = \left(1-R\right)\int^{z\rightarrow \infty}_{z_i}\dot{\rho_*}\left[(1+z)H(z)\right]^{-1}{\rm d}z .
\label{instant}
\end{equation}
For the SA models, $R = 0.4588$ is the fraction of mass returned to the ISM by SNe and stellar winds, which is a constant set by the adopted IMF. While in \gl~all the stellar mass is locked in galaxies, \lgl~models tidal stripping and thus a fraction of the stellar mass density is associated
with the intracluster light (ICL), rather than with a particular galaxy. The
exact fraction of stellar mass associated with the ICL in \lgl~depends on the
environment and formation histories of the galaxies involved, but we have found $\approx 20$\%  to be a good average approximation.

The
overall build-up of the comoving stellar mass density for the two SA models, \gl~and
\lgl,  is similar, as expected from the general agreement of their GSMFs
at $z=0$ and higher redshifts, seen in Fig.~\ref{fig:smf}, in particular around $\mb$. The stellar mass density in the \eg~model is lower than those from the two SA models: by a factor of 1.8 at
$z=0$ and a factor of 1.26 at $z=2$. The lower stellar mass density arises  from the lower normalisation of the \eg~GSMF at $M_* \sim \mb$ at $z=0$, seen in Fig.~\ref{fig:smf}. We have confirmed, by comparing $\rho_*$ in different stellar mass bins, that the difference in the stellar mass density is mainly due to galaxies with $M_*\ge
10^{10.5}~\msun$, around the break of the GSMF at $z=0$. By varying the parameters controlling different physical processes in \gl, we find that the stellar mass density evolution is strongly affected by the efficiency of feedback. In particular, the build-up process followed by galaxies with stellar masses
around and above $\mb$ at $z=0$ is strongly affected by the efficiency of AGN feedback \citep[see also][]{voort12,crain15}.

Fig.~\ref{fig:sm_vs_z} also shows the stellar mass density of star-forming galaxies
(dashed lines). At $z=0$, star-forming
galaxies in both \eg~and \gl~contribute $\approx 65$\% of the total stellar
mass density, while the contribution in \lgl~is $\approx 80$\%. These values depend on the definition used for selecting star-forming galaxies, however, very similar results have been found when changing the boundary value by $0.3$~dex. Thus, the larger fraction of star-forming galaxies might point to \lgl~having less efficient AGN feedback throughout cosmic time, given that the stellar mass density is dominated by galaxies with $M_*\approx M_{\rm Break}$. 

Note that a model that adopts an instantaneous stripping of the hot gas in
satellites has larger numbers of passive galaxies at $z=0$, and thus, builds its stellar mass more rapidly at $z>1$.

\subsection{The stellar mass-halo mass relation}\label{sec:shmass}

\begin{figure}
\includegraphics[width=3.0in]{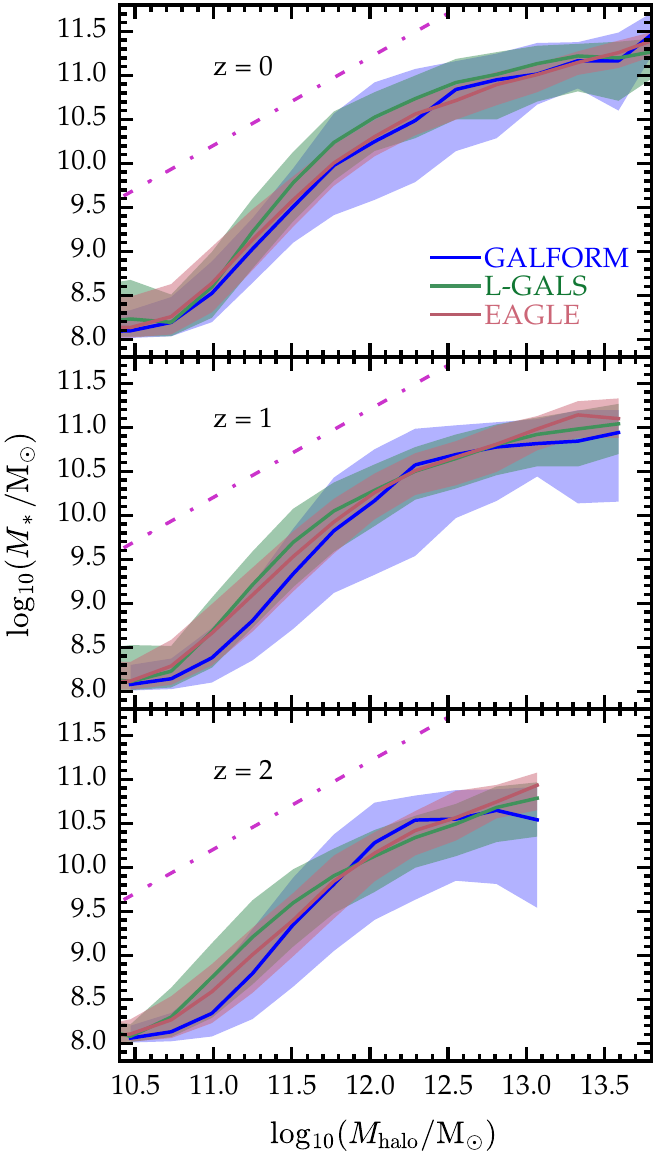}
\caption{\label{fig:halo_vs_sm} The stellar mass-halo mass relation ($M_*-
M_{\rm halo}$) for {\it central} galaxies with masses above $10^8\, M_{\odot}$, predicted
by \eg, \gl~and \lgl~at three redshifts, as indicated by the legend (see the
Appendix for a discussion of the halo mass definition).  The filled regions show the $10^{\rm th}$ and $90^{\rm th}$ percentiles of each
distribution, while the solid lines show the medians. The sloping cyan dash-dotted lines show the result of multiplying the halo mass by the universal baryonic fraction. At high masses, the stellar growth is driven by that stellar mass accreted during mergers, flattening the $M_*-
M_{\rm halo}$ relation. The scatter in this relation is larger for \gl~because (i) its AGN feedback depends more strongly on the accretion history of haloes and (ii) its stellar feedback efficiency is different in bulges and discs.}
\end{figure}

The $M_*-M_{\rm halo}$ relation, relation can be directly related to the star formation efficiency in haloes of different masses \citep[e.g.][]{Behroozi13}. Converting baryons into stars can be viewed as an inefficiency process, as an illustration of this point, the stellar masses of central galaxies obtained by multiplying the host halo masses by the universal baryonic fraction is shown as a dash-dotted line in Fig.~\ref{fig:halo_vs_sm}. 

\begin{figure*}

  \includegraphics[height=3.0in]{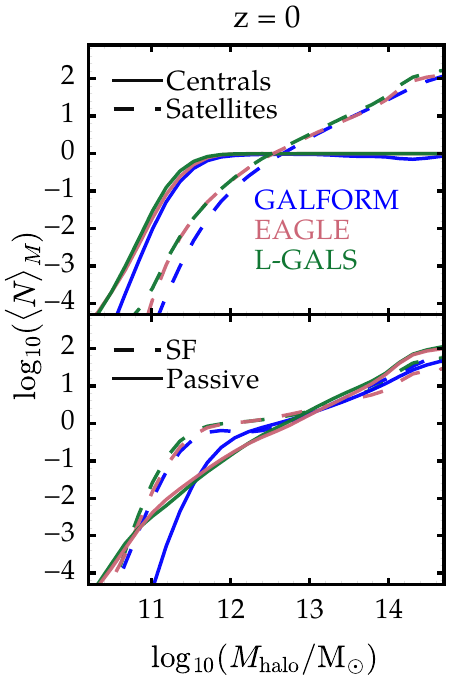}
  \includegraphics[height=3.0in]{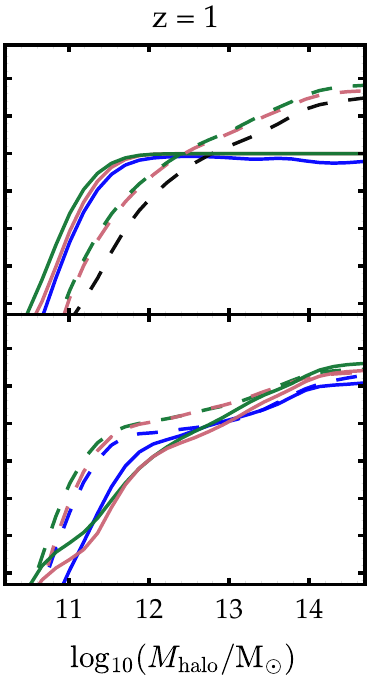}
  \includegraphics[height=3.0in]{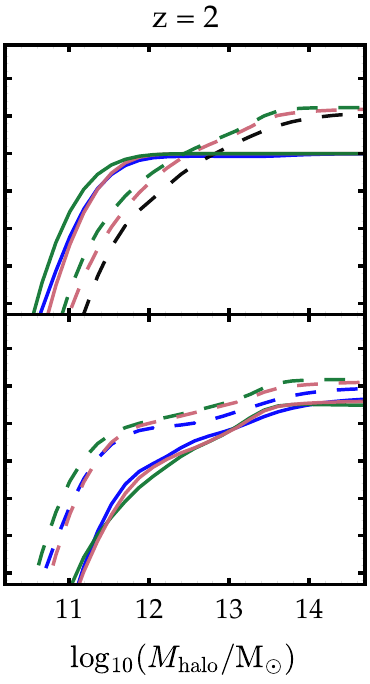}
 
  \caption{\label{fig:hod} The mean number of galaxies per halo, \nm,  of galaxies
  more massive than $M_{*} > 10^{9.5}~\msun$ in \eg, \gl~and \lgl~at
  different redshifts, from left to right: $z=0,1,2$. Top panels show
  the $\langle N\rangle_{M}$ of galaxies separated into centrals (solid lines) and
  satellites (dashed lines). Bottom panels show the $\langle N\rangle_{M}$ of
  galaxies separated into passive (solid lines) and star-forming (dashed lines, see Fig.~\ref{fig:ssfr_m} in \S\ref{sec:sf}). Differences in \nm~between the three models are below 0.5~dex, except for the minimum halo mass required to host passive galaxies. This larger difference is related to the population of ejected satellites discussed in \S\ref{sec:sf_def}.} 
\end{figure*}

In this section we explore the $M_*-M_{\rm halo}$ relation for {\it central} galaxies in \eg, \gl~and \lgl, as shown in Fig.~\ref{fig:halo_vs_sm}. $M_{\rm halo}$ is defined to be $\mth$, as described in the Appendix. The differences between the halo mass functions of the three models are negligible relative to the differences reported below in either Figs.~\ref{fig:halo_vs_sm} or~\ref{fig:hod}. Only galaxies with stellar masses $M_* > 10^8~\msun$ are studied here (see \S\ref{sec:sf_def}). 

At all redshifts, medians of the $M_*-M_{\rm halo}$ relation of the three models 
differ by less than 0.5~dex. The $M_*-M_{\rm halo}$ relations in Fig.~\ref{fig:halo_vs_sm} follow similar
trends for the three models at different redshifts: a monotonically increasing relation with
changes of slope at low, $\approx 10^{10.7}~\msun$, and high, $\approx
10^{12}~\msun$, halo mass. The $M_*-M_{\rm halo}$ relation flattens out at  $M_* < 10^{8.5}~\msun$ because no limit is imposed on the minimum halo mass. The change in slope at $\approx 10^{12}~\msun$ is mostly related to AGN feedback and the
cooling recipes \citep{bower12,crain15}. For all three models, AGN feedback becomes
effective for galaxies hosted by haloes with $\approx 10^{12}~\msun$. Up to the moment when a 
halo reaches $\approx 10^{12}~\msun$, the stellar mass
growth of the galaxy within the halo is mainly driven by the consumption of available gas, hence the feedback from SNe is very important \citep[e.g.][]{guo08}. However, the
AGN feedback prevents this growth channel from being effective, so mergers
become the main driver for any further growth in stellar mass, flattening the
$M_*-M_{\rm halo}$ relation. Given that the slope is shallower than unity, it is expected that minor mergers are the most important contributors to the stellar mass growth of massive galaxies. Even in the absence of AGN feedback, the stellar
mass growth is expected to be slower for very massive haloes due to the
increasing cooling times of massive haloes \citep[e.g.][]{silk77,cowie77,lu11cooling,mon14}. For \gl, we find
that in the absence of AGN feedback the change in the slope of the $M_*-M_{\rm halo}$ relation happens at $\approx 10^{12.5}~\msun$.

The normalization, slope and scatter of the median $M_*-M_{\rm halo}$ relations
shown in Fig.~\ref{fig:halo_vs_sm} mainly depend on the particular
implementation of the stellar and AGN feedback. Thus, as expected,
Fig.~\ref{fig:halo_vs_sm} shows differences between the models that vary with
the halo mass. At $z=0$, \lgl~has a steeper slope in the range $10^{10.75}<M_{\rm halo}/\msun<10^{12}$ than the other two
models. This gives rise to  more massive galaxies at a given host halo mass
above $10^{11}~\msun$. These aspects suggest that \lgl~has weaker AGN feedback
than the other models. 



The scatter in the $M_*-M_{\rm halo}$ relation for haloes in the \eg~simulation with $M_{\rm halo}\gtrsim10^{11}~\msun$, remains almost constant with redshift at $\approx 0.3$~dex. For haloes in \lgl~with $M_{\rm halo}\approx10^{11.5}~\msun$, the scatter of the $M_*-M_{\rm halo}$ relation decreases from $\approx 0.5$~dex at $z=2$ to $\approx 0.3$~dex at $z=0$, while it remains approximately constant $\approx 0.3$~dex for haloes with $M_{\rm halo}\approx10^{12}~\msun$. For haloes with $M_{\rm halo}\gtrsim10^{12}~\msun$, the scatter in the distribution
for the \gl~model is $\approx 0.8$~dex, and thus, larger than that for the other two
models. In this range of high masses there are two
aspects directly affecting the large scatter seen for the \gl~model. The first
one is that the efficiency of the AGN feedback in this model depends more
strongly on the accretion history of haloes, because the effect of AGN feedback explicitly depends on the ratio of the
halo cooling to dynamical time, taking into account the halo formation
history \citep{bow06,bower12}. The second aspect is related to \gl~having
different stellar feedback efficiencies in bulges and discs, as described in \S\ref{sec:samp}, while \lgl~assumes the SN mass loading depends on the halo maximum circular velocity,
which is naturally more tightly correlated with the mass of the host halo. This difference results in a greater scatter at $M_{\rm halo}\gtrsim10^{12}~\msun$ in the $M_*-M_{\rm halo}$ relation for \gl~than for \lgl~\citep{mit15}. Lowering the
efficiency of the stellar feedback reduces the scatter in
Fig.~\ref{fig:halo_vs_sm} for \gl, while increasing the normalization for small
haloes.

\subsubsection {The mean halo occupation distribution}\label{sec:hod}

The mean number of galaxies per halo, $\langle N \rangle_{M}$, which satisfy a particular selection criterion, as a function of halo mass is directly related to the
one-halo term for the clustering of those galaxies \citep[e.g.][]{benson00}. In
Fig. \ref{fig:hod}, we compare \nm~for galaxies with stellar mass $M_* >
10^{9.5}~\msun$ from the three models, at different redshifts. This stellar mass cut is
chosen so that we can further separate galaxies into star-forming and passive categories,
taking into account the resolution limits in \eg, as described in \S\ref{sec:sf_def}.

Galaxies with  $M_* > 10^{9.5}~\msun$  start to appear in haloes of minimum
mass $M_*\approx 10^{11}~\msun$, consistent with the stellar mass-halo mass relation
shown in Fig.~\ref{fig:halo_vs_sm}. The predicted \nm~for galaxies
chosen with different cuts in stellar mass follow similar trends to those shown
in Fig. \ref{fig:hod}. However, the higher the cut in stellar mass, the more
massive the host haloes are. 

The top panels in Fig. \ref{fig:hod} show the \nm~separated into central and
satellite galaxies, while the bottom panels show \nm~of
star-forming and passive galaxies (see \S\ref{sec:sf_def} for the definition of the
split). The \nm~for central galaxies is close to a unit step function, 
which indicates that for a massive enough halo we can
always find one central galaxy more massive than $M_* = 10^{9.5}~\msun$. This does
not hold if either the cut in stellar mass is set to $M_*\gtrsim
10^{10.5}~\msun$  or if additional cuts are applied, such as in colour. In both
cases, a decline in the \nm~for central galaxies is expected, since galaxies
selected in such a way will be less common \citep[e.g.][]{gp11,zehavi11}. 

As shown in Fig.~\ref{fig:hod}, the \nm~for satellite galaxies is basically a
power law beginning at halo masses about an order of magnitude larger than the
minimum mass required to host a central galaxy at a given redshift. The three
models predict comparable trends for both central and satellite galaxies. At $z=0$, the difference in numbers is less than $0.5$~dex. This difference between \gl and the other two models, increases to $\sim 0.8$
at $z=2$, for most of the mass range covered. 

The bottom left panel in Fig. \ref{fig:hod} shows that at $z=0$ and in haloes with masses $\gtrsim 10^{12.5}~\msun$ the contribution from
passive and star-forming galaxies to the global \nm~is comparable (values within $0.5$~dex). This is not the case when
an instantaneous stripping of the hot gas in satellites is assumed in \gl~(not shown). In
this case, most satellites become passive very quickly, which translates into a
\nm~dominated by passive galaxies also at the massive end. 

At $z=0$, passive galaxies in \gl~populate larger haloes than in either \lgl~or \eg. The same is true for central galaxies, although the difference is smaller. This is directly related to the different assignment of galaxies to haloes as was discussed in \S\ref{sec:sf_smf}. Changes in the parameters controlling the feedback in \gl~have an impact on the minimum halo mass for hosting a galaxy above a given mass. Thus, the small differences, $\approx 0.2$~dex, in the \nm~of star-forming galaxies
derived with the three models under study are at least partly driven by the differences
in the efficiency of the stellar feedback and also by the treatment of mergers, as was discussed by
\cite{contreras13}. 

\section{The cosmic star formation rate density}\label{sec:sfrd}

\begin{figure}
  \includegraphics[width=3.0in]{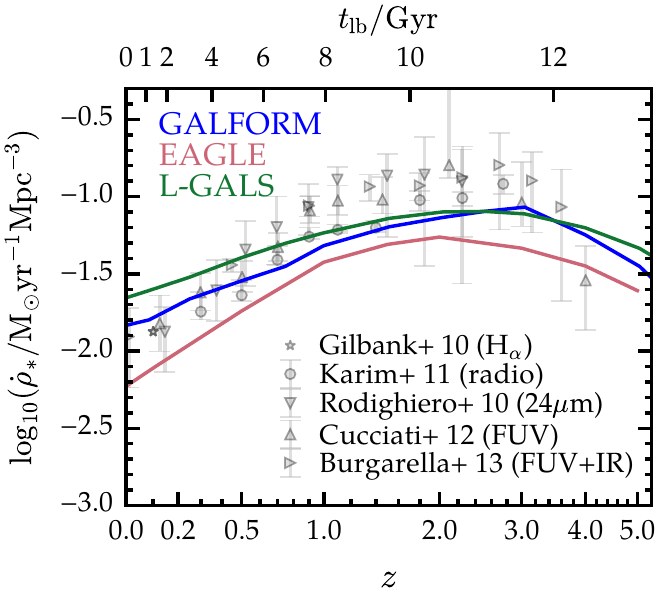}
  \caption{\label{fig:sfrd_vs_sm} The evolution of the cosmic star formation
  rate per unit comoving volume (SFRD) of galaxies with stellar masses
  $M_{*}>10^{8}~\msun$ predicted by \eg, \gl~and \lgl, as indicated by the
  legend.  For reference, the grey symbols present the observational results
  from \citet{gilbank2010}, which are based on H$\alpha$ measurements,
  \citet{rodighiero2010}, based on 24$\mu$m, the radio measurements from
  \citet{karim2011}, the FUV measurements from \citet{cucciati2012}, and the
  combined FUV$+$FIR measurements from \citet{burgarella2013}.  When necessary,
  the data have been converted to a \citeauthor{cah03} IMF and Planck Cosmology.
  The slope of the SFRD between $z=0$ and $z=1$ in \eg~is about a factor of
  $1.5$ steeper than those from the two SA models. This suggests that some of
  the physical processes that evolve with cosmic time in \eg~do not evolve as
 much in the two SA models.}
\end{figure}

The predicted evolution of the cosmic star
formation rate density (SFRD) is shown in Fig.~\ref{fig:sfrd_vs_sm} for galaxies with $M_{*}>10^{8}~\msun$, for the
three models (see section \S\ref{sec:sf_def} for a discussion on the stellar mass limits). We find that the SFRD is dominated by star-forming galaxies with
stellar masses around the knee of the GSMF, $M_{*}\approx 10^{10.5}~\msun$ (not shown). Thus, the contribution to the
global SFRD from galaxies with $M_{*}<10^{8}~\msun$ is negligible.

The general shape of the predicted SFRD
evolution is similar for the \eg, \gl~and \lgl~models, with a rise from $z=0$
to about $z\approx 2.5$ and a decline at higher redshifts, as shown in Fig.~\ref{fig:sfrd_vs_sm}. However there are important quantitative differences between the models.

The normalisation of the SFRD is different for the three models, with a difference at $z=0$ of a factor of $3.5$ between \eg~and \lgl~and of $2.5$ between \eg~and \gl. The difference in normalisation between \eg~and the two SA models is relatively persistent at $z<5.0$. The SFRD at $z=0$ differs by a factor of $1.5$ between \gl~and \lgl, converging at $1<z<3.5$. 

%
%

The peak of the SFRD is
predicted to occur later in the \eg~model, $z_{\rm peak}\approx 2$, than in
either \gl~or \lgl, $z_{\rm peak}\approx 3$. The slopes of the SFRD between $z=0$ and the corresponding $z_{\rm peak}$ are about a factor of $2$ flatter for the two SA models compared to \eg~(this factor is reduced to $\approx 1.5$ in the redshift range $0<z<1$). Thus, in \eg~the growth of the stellar mass is delayed with respect to the SA models, but then the star formation quenching happens faster.

We have not found a satisfactory explanation for this difference by varying the free parameters in \gl~that control either the gas cooling or the efficiency of the feedback processes. The slope of the SFRD evolution in \eg~is influenced by processes that evolve with time due to the changing characteristics of the ISM and the intergalactic medium. These processes, while naturally incorporated in \eg, need a explicit modelling in the two SA models. Some aspects that have been suggested to evolve with time are the cooling of gas in massive haloes \citep{dekel09}, the associated effect on the efficiency of AGN feedback \citep{voort12}, the stellar-driven outflows \citep{hopkins14} and the reincorporation timescale of gas \citep{mit14,henriques14}. On these last two points, \citet{hirschmann16} showed that the evolution in time of the gas reincorporation timescales is not captured in the SA models used in this work. This aspect is fundamental for the evolution of galaxies and a more detailed analysis of outflows should be made in comparison to observations, in order to find a model that captures more realistically the behaviour of gas flows.

\begin{figure*}
  \includegraphics[height=4.0in]{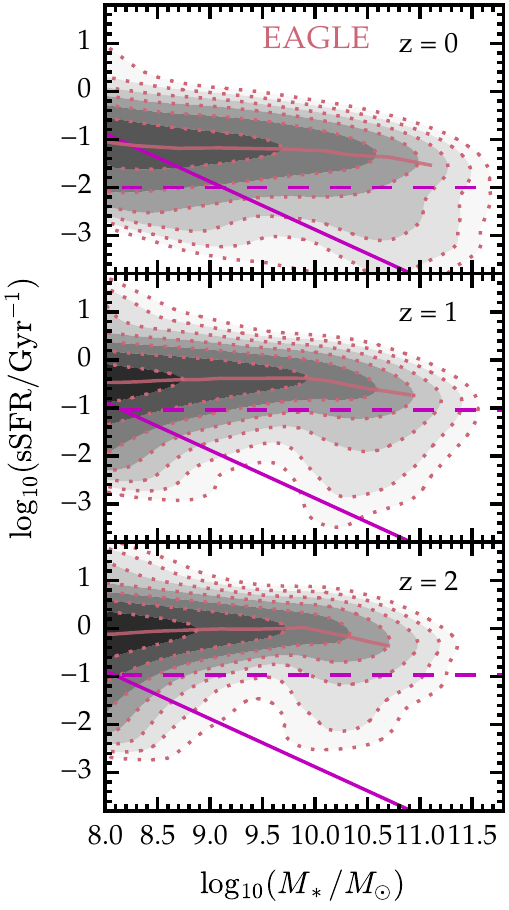}
  \includegraphics[height=4.0in]{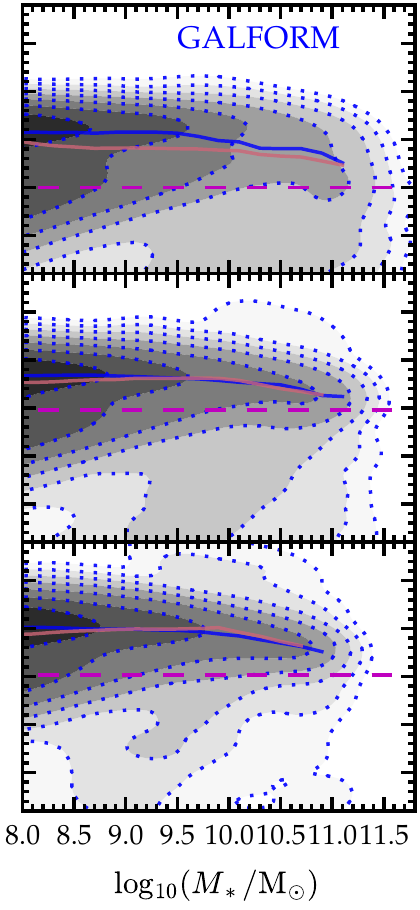}
  \includegraphics[height=4.07in]{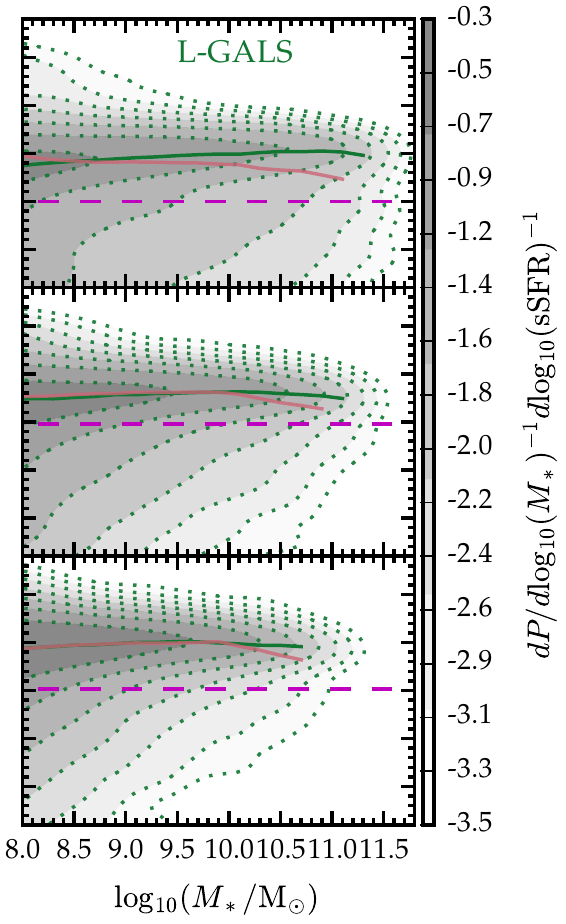}

 \caption{\label{fig:ssfr_m} The probability densities of all galaxies as a
 function of sSFR and stellar mass predicted by \eg, left column, \gl,
 middle column, and \lgl, right column, at three redshifts, as indicated
 by the legend. The contours correspond to number densities divided by the total number of galaxies in the sample. The cyan horizontal dashed lines separate galaxies
 into star-forming, above, and passive galaxies, below.  These lines are placed
 at $\log_{10}({\rm sSFR}/{\rm Gyr}^{-1}) = -2,\, -1.04,\, -0.97$ for $z=0,\, 1,\, 2$, respectively. The
 median sSFR of star-forming galaxies as a function of stellar mass is shown as
 a solid red line for \eg~(in all panels), a solid black line for \gl~and a solid
 green line for \lgl.  The cyan sloping solid lines are a guide to the resolution of
 the \eg~simulation, corresponding to the minimum stellar mass for galaxies with
 30 SF particles resolving the sSFR value (see \ref{sec:sf_def} for further details). Note that the intersection between the guide to \eg~resolution, cyan sloping solid lines, and the boundary between passive and star-forming galaxies, cyan horizontal dashed lines, evolves with redshift
 due to the evolving sSFR limit. The median sSFR of star-forming galaxies is remarkably similar for the three models, and the small variations shown are driven by differences in the feedback efficiency.}
\end{figure*}
\section{Specific star formation rates}\label{sec:sf}
\begin{figure*}
  \includegraphics[height=3.0in]{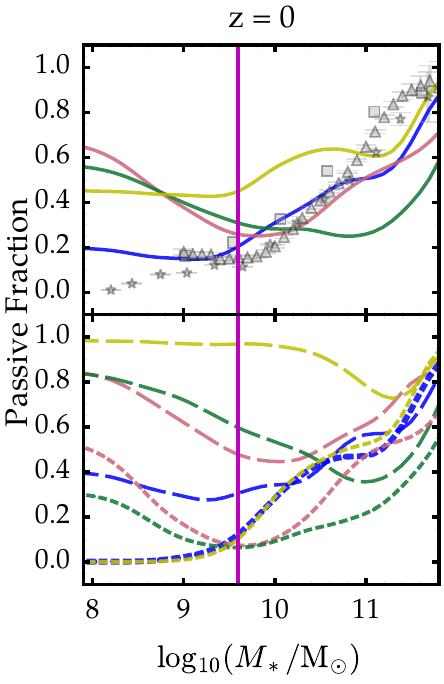}
  \includegraphics[height=3.0in]{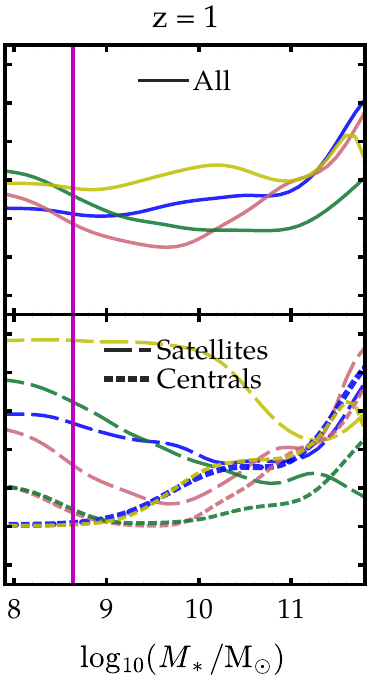}
  \includegraphics[height=3.0in]{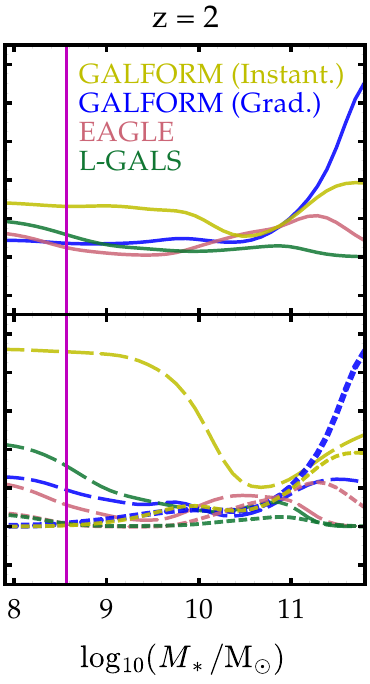}
  \caption{\label{fig:fraction_vs_sm} The fraction of passive galaxies as a function of stellar mass, predicted by
  \eg, \gl~and \lgl~at different redshifts, as indicated by the legend. The top
  panels shows the passive fractions for all galaxies, while the bottom
  panels shows separately the contributions from central (dotted lines) and
  satellite galaxies (dashed lines). In this plot we show the predictions from \gl~assuming either gradual ("Grad.",  black lines, the default model in this paper) or instantaneous ram pressure
  stripping ("Instant.", yellow lines) of the hot gas in satellite galaxies.  The cyan vertical lines are a guide to the
  resolution limits of \eg, note that these evolve with redshift due to an
  evolving sSFR limit (see \S\ref{sec:sf_def}). For reference, observational passive fractions based on either colour or sSFR cuts are plotted
  as grey stars \citep{gilbank2010}, triangles \citep{moustakas2013} and squares \citep{bauer2013}. The passive fraction increases
  with cosmic time. The increase in the fraction of low-mass passive galaxies in
  \eg~and \lgl~is driven, at least in part, by the population of ejected
  satellites discussed in \S\ref{sec:sf_smf}. 
}
\end{figure*}

The specific star formation rate (the star formation rate per unit stellar
mass in a galaxy, sSFR$=$SFR$/M_*$) gives a measure of the inverse of the time scale a galaxy requires to
assemble its stellar mass in situ by a constant star formation rate. The distribution of the predicted sSFR as a function of stellar mass for \eg,
\gl~and \lgl~are shown in Fig.~\ref{fig:ssfr_m}, at redshifts 0, 1 and 2. For all three models, this distribution
is dominated by galaxies with sSFRs between $\approx$0.1 and 10 Gyr$^{-1}$ across two orders of magnitude in stellar mass. A drop in the sSFR distribution at high masses is also seen in Fig.~\ref{fig:ssfr_m},
$M_*>10^{9.5}$M$_{\odot}$, for \gl~and \eg. This is related, at least in part, to the fact
that in this regime, the same mass haloes can host galaxies with very different
SFRs, depending on the activity history of their central black hole (Bower et
al., in prep.).

\subsection{The star-forming sequence}\label{sec:sf_seq}
The median sSFRs of star-forming galaxies from \eg, \lgl~and \gl~at $z=0,\,
1,\, 2$, are shown in Fig.~\ref{fig:ssfr_m} as solid lines. The scatter in the sSFR of star-forming galaxies at a given stellar mass is similar among the
three models, although the range is marginally larger for $M_*<10^9\msun$ in the \lgl~model. We have tested the impact of varying the sSFR cut by up to $1$~dex below the chosen boundary values (see \S\ref{sec:sf_def}) and it does
not change the global trends reported here for \lgl. 

The median sSFR for star-forming galaxies remains remarkably constant with
increasing stellar mass for all three models in the studied stellar mass range.
Previous comparisons with observations have shown that model star-forming
galaxies have a too flat median sSFR as a function of stellar mass
\citep[e.g.][]{wei12,mit14,fur14}. 

At $z=1$ and $2$ the models agree within $\approx 0.2$~dex. However, at $z=0$, \gl~predicts a median sSFR that is $\approx 0.4$~dex higher than that of \eg~and \lgl~for galaxies with $M_*<9.5~\msun$. In this stellar mass range, the GSMF in \gl~is dominated by star-forming galaxies, with number densities larger than for the other models (see Fig.~\ref{fig:fig5_smf_split}). This suggests that the stellar feedback in \gl~is weaker than in \eg~and \lgl, although, given that the difference is only found at $z=0$, a more detailed study should be made in order to understand the origin of this difference.

Further, at all redshifts and above $M_*>10^{10}~\msun$ there is a weak trend for the median sSFR to decrease with stellar mass for \eg~and \gl, which is not seen for \lgl. The median sSFR for star-forming galaxies predicted by the \lgl~model
is similar to those predicted by the other two models below
$10^{9.5}~\msun$ but the difference increases with stellar mass to a factor of $\sim 3$ for galaxies with
$M_*=10^{11}~\msun$ at $z=0$, being the largest at this redshift. Reducing the AGN feedback efficiency in \gl~increases the median sSFR for $M_*\gtrsim 10^{10}~\msun$ at $z=0$, while leaving it unchanged at lower stellar masses. Moreover, from Fig.~\ref{fig:fig5_smf_split} it is clear that at $z=0$ the abundance of star-forming central galaxies around the knee  of the GSMF, $M_*\approx10^{10.75}~\msun$, is larger for \lgl~than for the other two models. As shown in Fig.~\ref{fig:halo_vs_sm}, these galaxies are also hosted by less massive haloes in \lgl, compared with the other two models. Both aspects are directly connected with \lgl~having a weaker AGN feedback, allowing for a higher fraction of massive, $M_*>10^{10.5}~\msun$, star-forming central galaxies at $z=0$ which are hosted by haloes with $M_*\approx 10^{12}~\msun$. Thus, a weaker AGN feedback could explain the difference between \lgl~and the other models.

\subsection{The passive fraction}\label{sec:passive}

In both \eg~and \lgl~there are galaxies with SFR$=0$ and $M_*>10^{8}~\msun$
(over 25\% at $z=0$). The existence these galaxies is the
result of both the finite resolution of the simulation and the threshold in cold gas density imposed for the formation of stars (see \S\ref{sec:models} for further details).  At any redshift, less than 1\% of \gl~galaxies has SFR below $10^{-5}~\mathrm{Gyr^{-1}}$. Most of the
galaxies with low sSFR are star-forming discs, for which \gl~assumes the surface SFR density to be
proportional to the molecular surface density \citep{lagos11}. This aspect is critical since all
galaxies will convert some of their gas into $H_2$, resulting in a non-zero SFR as opposed to the modelling done in both \eg~and \lgl. 

Besides those galaxies with SFR$=0$, it can be seen in Fig.~\ref{fig:ssfr_m} that both \gl~and \lgl~predict a
relatively larger spread of passive galaxies with sSFR varying from 0.01 to
$10^{-5}~\mathrm{Gyr}^{-1}$. However, most passive galaxies in \eg~have sSFR
from 0.01 to $\approx 10^{-4}~\mathrm{Gyr}^{-1}$. This reduced range is due to the minimum SFR that
can be resolved in the \eg~simulation, which depends on the mass resolution and
the density threshold for star formation \citep{schaye15}. 

The fraction of passive galaxies across cosmic time results from the interplay of the different physical processes that quench star formation in galaxies, such stellar and AGN feedback and environmental processes like ram pressure stripping of hot gas. The fraction of passive galaxies as a function of stellar mass is shown in Fig.~\ref{fig:fraction_vs_sm} for \eg, \gl~and \lgl~at $z=0,\, 1$ and 2. The three models predict a global increase of the passive fraction with cosmic
time. 

For $z<2$ and $M_*\gtrsim 10^{10}~\msun$, the three models predict the passive
fraction to increase monotonically with stellar mass. At $z=2$, the same is true for \gl, while both \eg~and
\lgl~predict a rather flat fraction of passive galaxies that remains below 20\% for
the whole population, and below 40\% for satellites. The strength of the AGN feedback impacts the number of massive
passive galaxies and drives, at least
partially, the differences between models at the massive end. \gl~has the strongest AGN feedback of the three models, producing a larger fraction of passive and massive, $M_*\gtrsim 10^{11}~\msun$, galaxies, while \lgl~has the weakest AGN feedback and thus has the lowest fraction. The stellar feedback can also impact the slope of the passive fraction at the massive end. In the \gl~model, the gas mass heated by stellar feedback is directly dependent on the circular velocity of the disc, which is used as a proxy for the gravitational potential. However, for massive galaxies, the tight relation between the circular velocity in discs and their host haloes disappears, affecting the passive fraction slope for $M_*\gtrsim 10^{10}~\msun$ . 

The \eg~simulation predicts the passive fraction to have a minimum $M_*\sim 10^{10}~\msun$ at $z=0$ and $M_*\sim 10^{9.5}~\msun$ at $z=1$. The passive
fraction below these stellar masses is however strongly affected by sampling, due to the
SFR and feedback being quantised, which gives rise to numerical effects
\citep{fur14}.

At $z=0$, \gl~predicts a smaller number of low-mass passive galaxies relative to the other models. This is related to the higher number density of low-mass star-forming galaxies seen in Fig.~\ref{fig:fig5_smf_split} and, at least partly, a consequence of \gl~galaxies experiencing a less efficient stellar feedback than in the other models. This also results in lower fractions of passive satellite and central \gl~galaxies at $z=0$, in relation to the other two models.

The number of low-mass passive central galaxies is smaller in \gl~relative to the other models, as expected from the GSMF shown in Fig.~\ref{fig:fig5_smf_split}. Besides the different stellar feedback efficiency, this variation is also related to the ejected satellite galaxies being classified as either centrals or satellites, as described in \S\ref{sec:sf_smf}. As seen in Fig.~\ref{fig:fig5_smf_split}, there is an increase in low mass passive central
galaxies in \eg~and \lgl, relative to \gl, though this mainly happens below
the \eg~resolution limit.

For galaxies with $M_*<10^{10}\msun$, the fraction of passive satellites is larger than for central galaxies at all redshifts. The difference is reduced for higher stellar mass galaxies, with passive fractions being similar between central and satellite galaxies for $M_*>10^{11}\msun$ at $z=0$ and $z=1$, and at $z=2$ also, in the case of \lgl. At $z=2$, the fraction of passive and massive central galaxies is larger than for satellites in both \eg~and \gl. The high-mass end of the passive fraction is mostly shaped by the AGN feedback.

By default, in the \gl~model used for this study we are assuming a gradual stripping
of the hot gas in satellite galaxies. If instead an instantaneous stripping is
assumed, the passive fraction of satellites  rises to 100\%
below a threshold stellar mass which evolves to higher values at lower
redshifts, being around $10^{10}~\msun$ at $z=0$. This also affects the global
passive fraction predicted by \gl, which is then close to 40\% for low-mass
galaxies at $z=0$, instead of the 20\%, shown in
Fig.~\ref{fig:fraction_vs_sm}.

\subsection{The evolution of the sSFR}\label{sec:ev_ssfr}
\begin{figure}
  \includegraphics[width=3.0in]{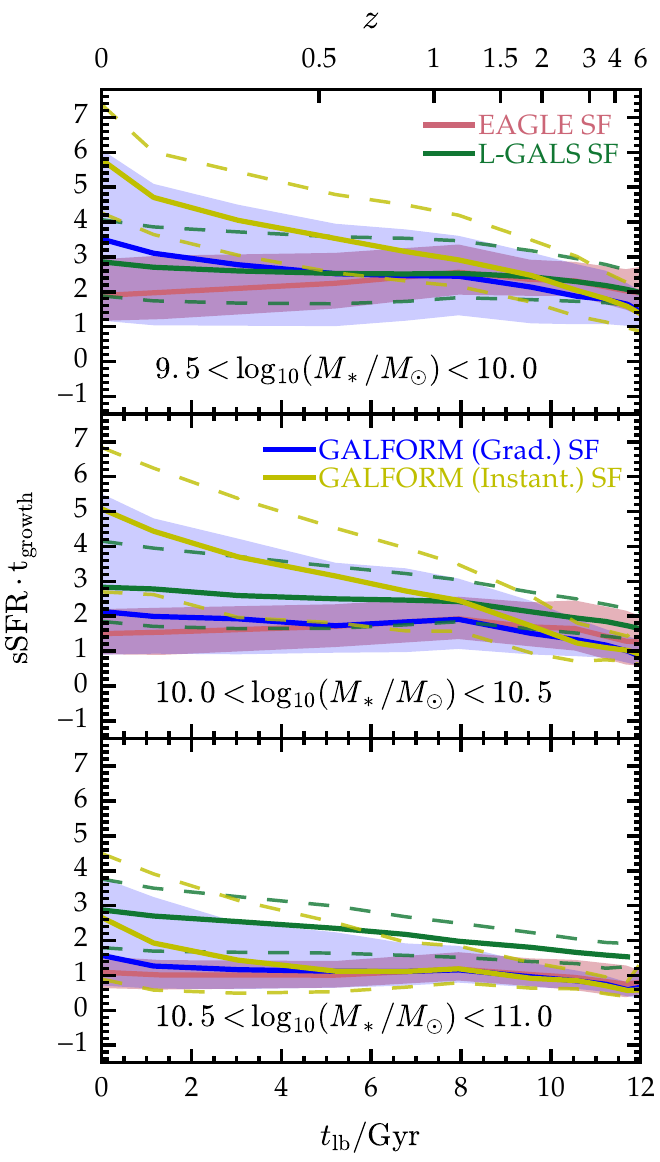}
  \caption{\label{fig:ssfr_vs_time} The median specific star formation rate times the expected growth timescale of the corresponding host haloes, 
   as functions of lookback time for star-forming
  galaxies in \eg, \gl~and \lgl~(see the text in \S\ref{sec:ev_ssfr} for further details). The three panels show star-forming galaxies in
  different stellar mass bins of 0.5~dex width and centred at
  $\log_{10}(M_{*}/\msun)=9.75, 10.25, 10.75$, as indicated by the legend.  The shadowed
  regions and dashed lines show the corresponding $25^{\rm th}$ and $75^{\rm th}$ percentiles of the distributions. The star-forming galaxies are classified
  according to the cyan horizontal lines in Fig.~\ref{fig:ssfr_m}, which are
  interpolated to other redshifts. In this plot we show the predictions from \gl~assuming either gradual ("Grad.", black lines, the default model in this paper) or
  instantaneous ram pressure stripping ("Instant.", yellow lines) of the hot gas in
  satellite galaxies. Given the flatness of the relations shown in this figure, we can conclude that the evolution of the sSFR, in all the default models, closely follows the specific mass assembly history of their host dark matter haloes.} 
\end{figure}

%
%

In Fig.~\ref{fig:ssfr_vs_time} we explore the evolution of the median sSFR with
cosmic time. Previous studies using SA models have shown that the stellar mass growth for galaxies of a given final mass
roughly follows that of the dark matter component but with a normalization that varies with mass \citep{mit14}. From N-body simulations of $\Lambda$CDM cosmologies, the dark matter halo mass has been found to grow with redshift roughly as a power law in scale factor and have little dependence on halo mass \citep{wechsler02,genel08,fakhouri10}. Recently, \citet{Correa2015} derived the halo mass accretion history from the growth rate of initial density perturbations, by using the extended Press-Schechter approach \citep[e.g.][]{lacey93,neistein06}. In Fig.~\ref{fig:ssfr_vs_time} we show the
median sSFR of star-forming galaxies normalised by the specific accretion rate of the typical haloes hosting the galaxies of interest, $t_{\rm growth}^{-1}$. This specific accretion rate is calculated from:
\begin{equation}
t_{\rm growth}^{-1} = \frac{{\rm d}M/{\rm d}t}{M(z)},
\end{equation}
\begin{equation}
M(z) = M_0(1+z)^{\alpha}e^{\beta z},
\end{equation}
where $M_{0}$ is the final halo mass and both ${\alpha}$ and ${\beta}$ are functions of $M_{0}$ \citep[see][for further details]{Correa2015}. We note that, although we use the analytical scaling equations presented in \citet{Correa2015}, similar trends and conclusions are reached when using the approximations  provided by either \citet{genel08} or \citet{fakhouri10}.

In Fig.~\ref{fig:ssfr_vs_time}, we separated galaxies into three stellar mass ranges. We are interested in comparing the star formation rates of galaxies in each of these ranges with the expected mass assembly rate of their typical host halo. Thus, here we set $M_{0}$ at each redshift to be equal to the average mass of the haloes hosting galaxies with stellar masses within each bin shown in Fig.~\ref{fig:ssfr_vs_time}. Although some bias can be introduced by doing this, for the purpose of providing a rough estimate of how closely the sSFR follows the evolution of< the mass accretion rate of dark matter haloes, this is a reasonable approximation. In fact, to explore in detail the build up of the mass for galaxies in different final stellar mass bins, individual haloes should be followed across time \citep{mit14}.

The first thing to note from Fig.~\ref{fig:ssfr_vs_time} is that the predicted
sSFR$\cdot t_{\rm growth}$ is reasonably flat for all the default models, with
slopes $<\pm 0.4$Gyr$^{-1}$. This implies that the star formation histories of model
galaxies closely follow the mean mass assembly rate expected from their dark matter host haloes and that the differences between the models are minimal. In the SA models,
the stellar mass assembly process broadly traces that of haloes because the
mass loading 
and reincorporation efficiencies do not evolve significantly over the history of the galaxy for
typical star-forming galaxies at $z<1$ \citep[see][for details]{mit14}. What
might be surprising, is that galaxies in \eg~also follow the dark
matter assembly quite closely. In \eg, the mass loading for SN feedback does not depend explicitly on dark
matter or halo properties \citep{schaye15}, nevertheless the reincorporation time for reheated
gas is expected to depend on the halo properties.

The sSFR tends to trace the dark matter assembly as a function of lookback time
even for a wide range of variants of \gl~\citep{mit14}. Nevertheless, we
 have found a very large difference for the \gl~prediction when
assuming the stripping of hot gas in satellite galaxies to be instantaneous as opposed to of gradual (the default here), as seen in Fig.~\ref{fig:ssfr_vs_time}. The figure only shows star-forming galaxies, which in the case of the instantaneous
stripping model, will be dominated by central galaxies given that in this model satellite galaxies are mostly passive. The scatter seen for the default \gl~model is largely due to the contribution of star-forming satellite galaxies. These dominate the low stellar mass ranges at low redshift, as seen in Fig.~\ref{fig:fig5_smf_split}. It is clear from this that the evolution of the stellar feedback in \gl~is different from the other models, allowing for larger numbers of low-mass galaxies by adopting either a weaker stellar feedback or shorter reincorporation times.

\begin{figure*}

  \includegraphics[height=4.0in]{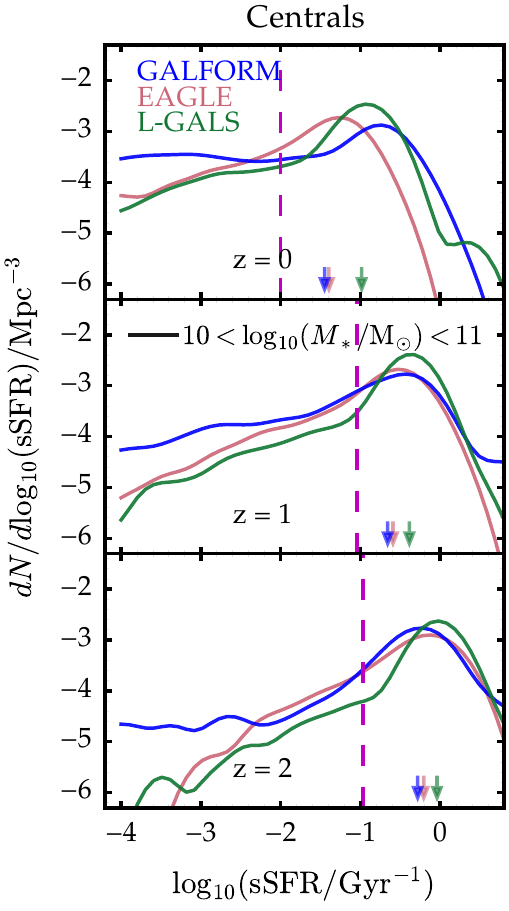}
  \includegraphics[height=4.0in]{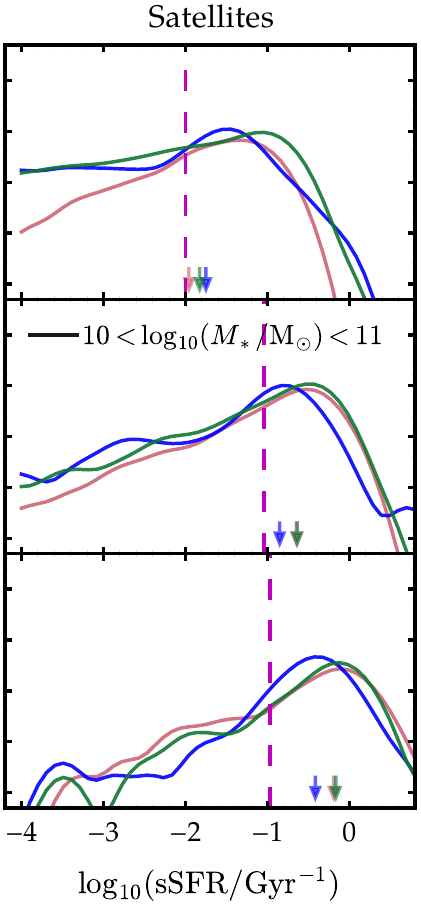}
  \includegraphics[height=4.0in]{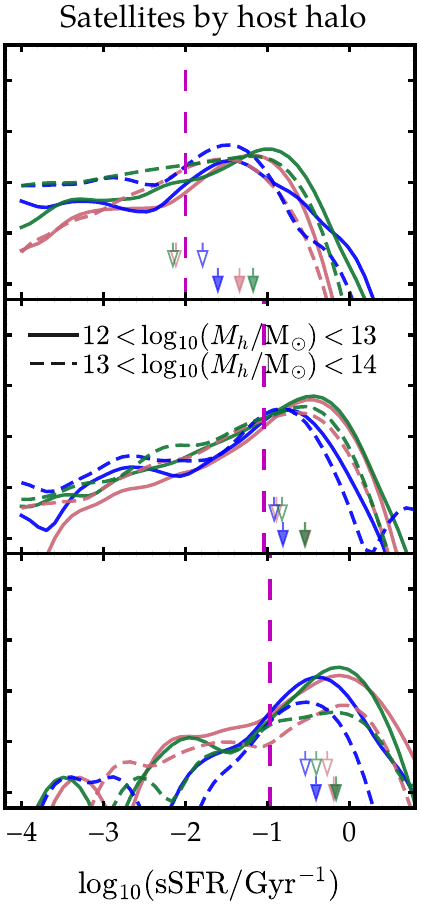}

  \caption{\label{fig:ssfr_hist_cen} The comoving density of central, left column,
  and satellite galaxies, central column, as a function of sSFR, for galaxies
  with $10^{10}<M_*/\msun< 10^{11}$, for the three models at three redshifts,
  as indicated by the legend.  The right column shows the number density of
  satellite galaxies with $10^{10}< M_*/\msun < 10^{11}$, hosted by haloes with
  $10^{12} < M_{\rm halo}/\msun<10^{13}$ (solid lines) and by haloes with $10^{13} <
  M_{\rm halo}/\msun < 10^{14}$ (dashed lines). In all the panels, the median sSFR for
  each model is shown by an arrow of the corresponding colour. In the right
  column, the medians of the lower mass ranges are shown by filled arrows, while
  the high mass ranges are shown by open ones. The dashed cyan vertical lines
  correspond to the chosen boundary between passive, left side, and star-forming
  galaxies, right side of the line (see also Fig.~\ref{fig:ssfr_m} and \S\ref{sec:sf_def}). Note from the right column of this figure, that the median sSFR of satellite galaxies with $10^{10}<M_*/\msun< 10^{11}$ declines with increasing host halo mass.}
\end{figure*} 
\subsection{The distribution of sSFR at a given stellar mass}\label{sec:ssfr}
Most of the stellar mass in the Universe at a given time is dominated by
galaxies located around the knee of the GSMF ($M_* \approx \mb$). Thus, it
is interesting to explore in more detail the distribution of sSFRs for these galaxies.  As has been discussed in \S\ref{sec:smf}, the stellar mass corresponding to the knee of
the GSMF depends on the model and, for a given
model, it evolves with redshift. In order to approximately enclose the relevant
range for the three models and redshifts, galaxies with $10^{10}<M_*/\msun<
10^{11}$ are included in Fig.~\ref{fig:ssfr_hist_cen}, separated into central and
satellite galaxies. In this stellar mass range, the three models predict similar number
densities at $z=0$ and there are sufficient satellite galaxies to study the trends even at higher redshifts. Note that while we focus on $10^{10}<M_*/\msun<
10^{11}$ here, the global trends and qualitative results are valid for stellar mass
ranges selected from about $10^{9.3}~\msun$ to $10^{11.5}~\msun$.

The three models present qualitatively
similar distributions for both central and satellite galaxies (see Fig.~\ref{fig:ssfr_hist_cen}), with galaxy
numbers increasing with sSFR until a peak is reached at sSFR between 0.03 and
1 Gyr$^{-1}$. Beyond which, the number of galaxies declines. For the mass range shown in Fig.~\ref{fig:ssfr_hist_cen}, the number of central
galaxies with low sSFR in \gl~is higher than for the other two models (over $\approx 1$~dex for galaxies with sSFR$<10^{-3.5}$Gyr$^{-1}$). This difference is
related to both the lack of a threshold for the star formation in this
model (see \S\ref{sec:passive}) and the details of the separation between
central and satellite galaxies, which affects the number of passive centrals, as
discussed in \S\ref{sec:sf_smf}.

The peak of the sSFR distribution occurs at somewhat different values depending on the
model. The sSFR medians increase with increasing redshift up to $z=2$ (see Fig.~\ref{fig:ssfr_hist_cen}), as the model SFRD does (see Fig.~\ref{fig:sfrd_vs_sm}). The sSFR medians for central galaxies in \lgl~is higher than
that in the other two models at all redshifts for this stellar mass  range, as expected from the difference seen for the median sSFR of massive galaxies (see Fig.~\ref{fig:ssfr_m}). 

\subsubsection{The sSFR as a function of environment}\label{sec:envir}
The environment can be traced by different estimators, such as the host halo mass,
kinematic parameters \citep{Hahn2007,Hoffman2012,Libeskind2012}, or the density
\citep{Sousbie2011,Tempel2011}. Here we use the host halo mass as
the tracer of the environment and thus, we note that the results might change if a different definition is chosen.  In the right column of
Fig.~\ref{fig:ssfr_hist_cen}, we compare the effect that environment has on the
sSFR of satellite galaxies as predicted by the three models.

We study satellite galaxies in the same range in stellar masses as in the middle
column of Fig.~\ref{fig:ssfr_hist_cen}, $10^{10}< M_*/\msun < 10^{11}$, separating them into halo mass bins of $10^{12}<
M_{\rm halo}/\msun < 10^{13}$ and $10^{13}< M_{\rm halo}/~\msun < 10^{14}$.  Modifying the stellar mass range of the
satellites, choosing mass ranges within $10^{9.3}<M_*/\msun<10^{11}$, again does not
change the results in this section. The same is true for the host halo mass,
provided that the chosen haloes are massive enough to host the corresponding
satellite galaxies.

Overall, the three models predict similar trends with environment (host halo mass, in this case). The
differences between the models on average are less than 1~dex.  For a given
stellar mass range, satellite galaxies hosted by less massive haloes have
slightly higher sSFRs than those hosted in more massive haloes. Correspondingly, at $z=0$
the fraction of passive satellite galaxies is higher for more massive host
haloes, though this trend is noisy due to the low numbers of passive galaxies.
For the two host halo mass ranges, the stellar mass distributions differ by less
than a factor of 3. Thus, the trends with environment seen for the sSFR of
satellite galaxies are not driven by differences in stellar mass but are a
reflection of the effect of environment (host halo mass) on  the star formation in the three
models.

\section{Metallicity}\label{sec:Z}

\begin{figure}
  \includegraphics[width=3.0in]{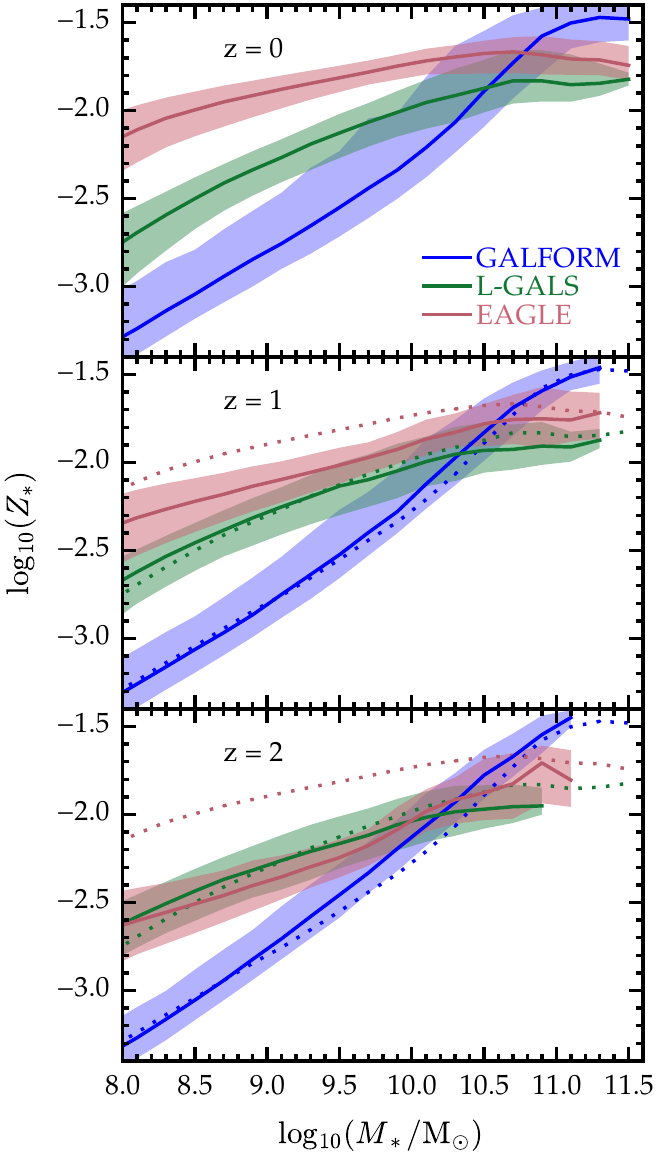}
  \caption{\label{fig:starz_vs_sm} The stellar metallicity as a function of stellar mass, the $M_*-Z_*$ relation, of galaxies in \gl, \eg~and \lgl, at different redshifts, as
  indicated by the legend.  The shaded regions show the
  corresponding $10^{\rm th}$ and $90^{\rm th}$ percentiles of the distribution.
  For reference, the median of the $M_*-Z_*$ relation at $z=0$ of each model is
  shown by dotted lines of the corresponding colour in the panels for $z=1$ and
  $z=2$. Note that the $M_*-Z_*$ relation from \eg\ is not converged relative to
  higher resolution simulations \citep{schaye15}. The slope of the $M_*-Z_*$
  relation from \eg~flattens with time and this is seen for both the reference
  run and the high resolution one, which is not shown in the figure. Meanwhile
  the slopes of the $M_*-Z_*$ relation from two SA models remain practically
  unchanged with redshift.}

\end{figure}

\begin{figure}
  \includegraphics[width=3.0in]{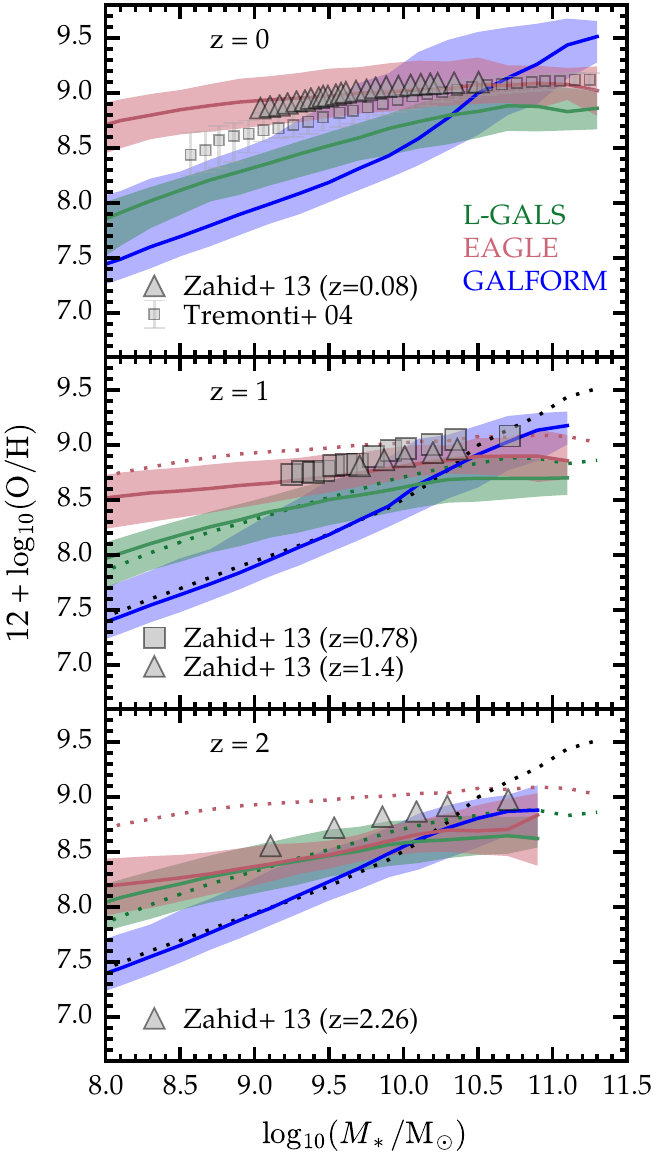}
  \caption{\label{fig:z_vs_sm} The cold gas metallicity (gas-phase oxgen
  abundance) as a function of stellar mass of star-forming galaxies, the
  $M_*-Z_{\rm cold}$ relation, in \eg, \gl~and \lgl, at different redshifts, as
  indicated by the legend.  The shaded regions show the corresponding $10^{\rm
  th}$ and $90^{\rm th}$ percentiles of the distribution. The median of
  $M_*-Z_{\rm cold}$ relation at $z=0$ of each model is shown by dotted lines in
  the $z=1$ and $z=2$ panels. For reference, values derived from observations by
  \citet{tremonti04} and \citet{Zahid13} have been included as grey symbols, as
  described in the legend. The $M_*-Z_{\rm cold}$ relations follow similar
  trends to the $M_*-Z_*$ ones, but with larger spreads.}

\end{figure}

The metallicity reflects the history of the gas reprocessing by stars and the
exchange of gas between a galaxy and its environment. Moreover, the metallicity
affects the cooling of the gas, which is one of the fundamental aspects driving
the efficiency and timing of star formation in galaxies. In this section, we
compare predicted metallicities, defined as the ratio between the mass of metals and
total mass. In the case of \eg, metallicities have been obtained by considering
either all particles associated with a given galaxy or those that are star-forming for the gas phase, as opposed to the aperture values applied to other properties.

An assumption that is made in the case of the SA models but not \eg, is the instantaneous recycling approximation, whereby all metals are returned immediately to the ISM following star formation. In \eg~ metals are returned to the ISM with a time delay which depends on their production time during stellar evolution. However, as the total $Z_*$ and $Z_{\rm cold}$ are dominated by oxygen, the majority of which is released on short timescales, this approximation is not bad at high redshifts, although the delayed contribution of intermediate mass stars is still neglected and this accounts for about half the metals at $z=0$ \citep{Segers2015}.

\subsection{The stellar mass-stellar metallicity relation}
\label{sec:Zstar}
%

The stellar mass - stellar metallicity, $M_*-Z_*$, relations of galaxies from the three  models at redshifts $z=0,1,2$ are shown in Fig.~\ref{fig:starz_vs_sm}. 
The models predict a stellar
metallicity that increases with stellar mass at all redshift shown. At $z=0$, the $M_*-Z_*$ relation flattens for $M_*>10^{10.5}~\msun$. For galaxies of $M_*\approx 10^{10.5}~\msun$ the median metallicities predicted by the three models are in marginal agreement (within $\approx 0.2$~dex) at all redshifts. Nevertheless, across the full stellar mass range, the stellar metallicities predicted by the models vary significantly.  At the range of redshift explored, the predicted slopes of the $M_*-Z_*$ relations are different between the 3 models. The slope of the $M_*-Z_*$ relation remains almost constant with redshift for both SA models, a slope of $\approx 0.6$ for \gl~and of $0.25$ for \lgl. The slope of the $M_*-Z_*$ relation varies in \eg~from $0.32$ at $z=2$ to $0.14$ at $z=0$. The difference in slopes between the models produces a large variation in the median metallicities of low-mass galaxies. This variation is largest at $z=0$. At this redshift, the median metallicity of galaxies with $M_* \approx 10^{8}~\msun$ differs by $\approx 0.5$~dex between the \eg~and \lgl~and by $\approx 1.1$~dex between \eg~and \gl. Similar
trends are found when we separate the sample into central and satellite galaxies. 

Both the slope and the normalisation of the $M_*-Z_*$ relation are sensitive to
the modelling of stellar feedback, and in the case of \eg~for $M_*\gtrsim10^{10}~\msun$, AGN feedback \citep{schaye15,Segers2015,lacey15}. Due to the very different ways in which hydrodynamic simulations and SA models treat metal enrichment and the transport of metals (\S\ref{sec:eg} and \ref{sec:samp}), these differences in the relations are unsurprising. In the two SA models, the stellar feedback can be reduced by either lowering the normalisation of the mass loading factor or by changing how it depends on the relevant velocity, as described 
in \S\ref{sec:samp}. Lowering the normalisation of the mass loading factor results in an increased normalisation of the $M_*-Z_*$ relation \citep{lacey15}, as is also found in \eg~\citep{crain15}. However, a  
 decrease in the slope of the dependence of mass loading on circular velocity
 allows small galaxies to retain a larger fraction of their metals, flattening
 out the $M_*-Z_*$ relation \citep{lacey15}. In \gl, the AGN feedback has a
 marginal effect on the $M_*-Z_*$ relation, while in \eg\ it can affect the
 massive end of this relation \citep{Segers2015}. At least at $z=0$, the
 differences between \eg\ and the SA models are alleviated when the latter
 include a non-instantaneous approximation for the recycling of metals
 \citep{yates13} and a higher resolution hydrodynamical simulation is used
 \citep{schaye15}.

An interesting difference between \eg~and the SA models is the lack of evolution in the slope of the $M_*-Z_*$ relation in the two SA models. This lack of evolution implies that, on average, the growth in
metal mass follows almost exactly the growth in stellar mass, while this is not
the case for \eg~galaxies. In SA models the differential recycling and metal retention effects are not properly accounted for \citep{Ma2015}. Stellar winds can affect the efficiency of stellar feedback and, in turn, the evolution of the metal retention in small galaxies. Winds in the SA models could be treated in a more realistic way, for example by using a dynamical
model of supernova feedback following the evolution of pressurised bubbles
\citep{lagos13}. We have tested that the pre-enrichment of the gas in haloes at late times has a minimal impact on the evolution of the $M_*-Z_*$ relation. One aspect that may affects the evolution of this relation is the delayed metal enrichment that occurs in \eg, as opposed to the instantaneous recycling approximation assumed in the two SA models in this work, as the stellar feedback depends on the metal enrichment \citep{yates13,delucia14}. 
 
The treatment of metals in \eg~is very different to that in SA models as the metals are associated with particles. Gas particles are enriched in a stochastic manner when in the vicinity of stars (\S\ref{sec:eg}), thus as winds are driven from the ISM the metals in the galaxy are redistributed. To partially account for the mixing of metals in the simulation, smoothed metallicities are computed using the SPH kernel of a particle \citep{Wiersma2009b}, however metal diffusion is not accounted for in the simulation. While aspects of the subgrid physics can be improved to account for the transport of metals, \cite{schaye15} have shown that the $M_*-Z_*$ relation is not converged relative to higher resolution simulations. Such higher resolution simulations produce a steeper relation, at least at $z=0$.
Thus, to carry out a more detailed study of the metallicities relative to hydrodynamical simulations, converged results are required first.


\subsection{The stellar mass-cold gas metallicity relation}
%
In Fig.~\ref{fig:z_vs_sm} we show the predicted stellar mass - cold gas
metallicity, $M_*-Z_{\rm cold}$, relation for
the three models at redshifts $z=0, 1$ and $2$. In particular we show the gas-phase oxygen abundance for star forming galaxies as this quantity is only measurable observationally for
such galaxies \citep[e.g.][]{tremonti04,Zahid13,zahid14}. The model $Z_{\rm cold}$ is converted to the gas-phase oxgen abundance shown in Fig.~\ref{fig:z_vs_sm} assuming solar abundance ratios. Specifically, we assume that $Z_{\msun}=0.0134$ \citep{Asplund2009} and $12+\log_{10}{\rm (O/H)_{\msun}} =8.69$ \citep{AllendePrieto2001}.
This selection of galaxies on or above the main sequence of
star formation also ensures that their properties are well resolved in \eg. The gas phase metallicity
depends on the flows of gas in and out of a galaxy and on the stellar winds from
dying stars and is thus expected to roughly follow the stellar metallicity but
with more variability. 

The models predict a gas phase metallicity that increases with stellar mass for
star-forming galaxies in a similar way to that shown for the stellar metallicity
in Fig.~\ref{fig:starz_vs_sm}, although with a larger scatter, as expected. As
for the stellar metallicity, we find little evolution for the gas phase
metallicity in the SA models, while \eg~predicts a relation that flattens with
increasing time and a small, $\approx0.2$~dex, increase in the mean metallicity. 

The reasons that account for these differences are similar to those discussed in
\S\ref{sec:Zstar}. In summary, the metallicities of simulated galaxies are found
to be inconsistent across models, both as a function of stellar mass and
redshift.  To investigate these differences further, higher-resolution
hydrodynamic simulations are required and more detailed modelling of metal
production and redistribution should be considered in SA models.

\section{Galaxy Sizes}\label{sec:r50}

\begin{figure}
  \includegraphics[width=3.0in]{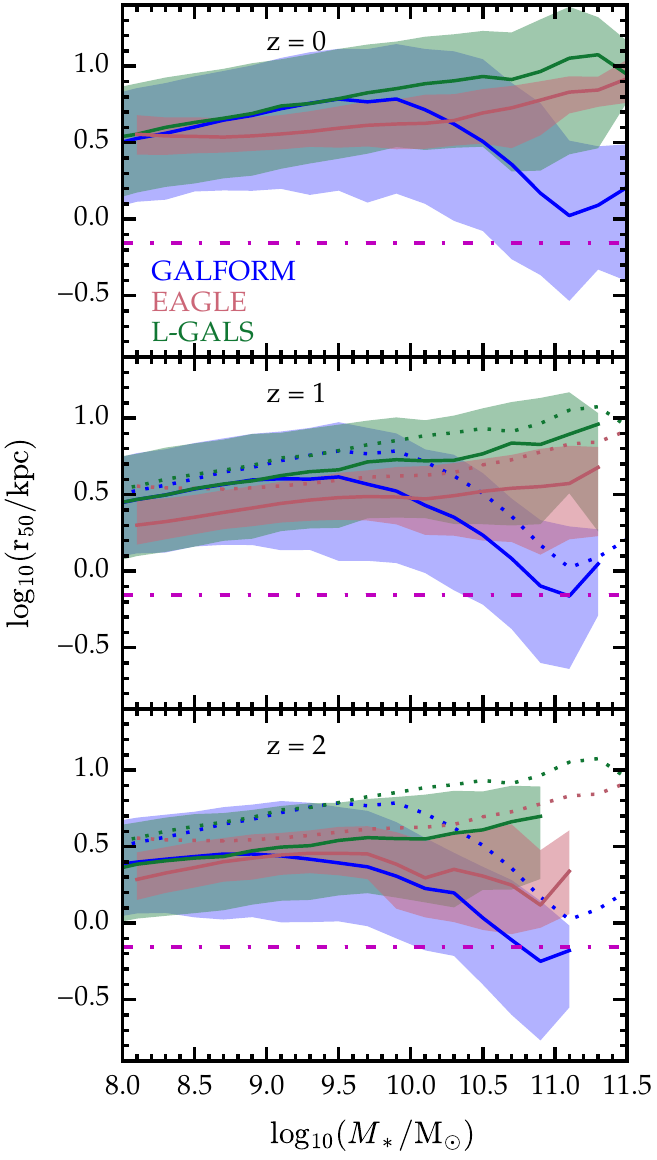}
  \caption{\label{fig:r50}
  The median half-mass radius ($r_{50}$) as a function of stellar mass, the $M_{*}-r_{50}$ relation, for
  galaxies in \eg, \gl~and \lgl~at different redshifts, as indicated by the
  legend. The shaded areas comprise the $10^{\rm th}$ to $90^{\rm th}$ percentiles of
  the distributions. The cyan  horizontal dot-dash lines indicate the scale of the
  gravitational softening used in the \eg~simulation at different redshifts, which can
  be used as a resolution guide. For comparison, the median $M_{*}-r_{50}$ relation at $z=0$ for each model is also shown in the $z=1$ and $z=2$ panels, as dotted lines of the corresponding colour. At low-masses, the scatter in the $M_{*}-r_{50}$ relation is very similar for the two SA models and about a factor of two larger than that for \eg. At high masses the \gl~$M_{*}-r_{50}$ relation drops due to the modelling of the contraction of dark matter haloes caused by the self-gravity of baryons.}
\end{figure}

In this section, we compare predicted galaxy sizes as measured by the 3D stellar
half-mass radius, $r_{50}$. This quantity, although not available
observationally\footnote{For \eg, the
projected half-mass radii for only disc galaxies (S\'ersic index, $n_s <
2.5$), which are calculated by fitting a S\'ersic profile to projected and
azimuthally averaged surface density profiles , are compared with the
observations in Fig. 9 of \cite{schaye15}. For \gl, the predicted half-light
radii as a function of r-band magnitude for early and late-type galaxies are
compared with the observations in Fig. A3 of \cite{gp14}. For \lgl, the
projected half-light radii for late-type galaxies are compared with observations
in Fig. 2 of \cite{qi11}.}, is the most straightforward way to encapsulate the galaxy size
and allows a fair and detailed comparison between the studied
models. 

Details on the calculation of
galaxy sizes by the two SA models can be found in \S\ref{sec:samp}. In \eg, the $r_{50}$ radius has been measured using a 3D aperture\footnote{See \cite{Furlong2015} for a discussion of the change in galaxy sizes with aperture radii in \eg.} of 30 proper kpc, as applied for the
stellar mass and SFR measurements discussed previously. 

In Fig.~\ref{fig:r50} we show the stellar mass-size, $M_{*}-r_{50}$, relation at $z=0,\, 1,\, 2$ from \eg, \gl~and \lgl.  
The dot-dashed horizontal lines indicate the gravitational force softening in \eg~(see Table~\ref{tbl:eagle}), below which galaxy sizes can be overestimated due to the suppression of the gravitational force at this limit.
Note that this does not affect galaxies presented in this study across the redshift range considered. 
A point to recall before discussing the comparison between models is that in \eg\ the sizes of disc galaxies at $z\approx 0$ were considered in the calibration of the model (Table~\ref{tbl:cali}), however the evolution of the galaxy sizes is a prediction from the model \citep[see][for a detailed study of the evolution of the galaxy sizes compared with observations]{Furlong2015}. 

%
The median $M_{*}-r_{50}$ relations for the three models at $z=0$ and $z=1$ increase with stellar mass for $M_*\lesssim 10^{9.5}~\msun$, while \gl~sizes decrease with increasing mass at $M_*\gtrsim 10^{9.5}~\msun$. 
At $z=2$, \lgl~sizes increase with stellar mass, the \eg~relation flattens at $M_*\approx 10^{9.5}~\msun$ and decreases with stellar mass at higher masses, and the \gl~sizes are reasonably flat with stellar mass at $M_*\lesssim 10^{9.5}~\msun$ and again decrease with increasing stellar mass at higher masses.
In spite of differing trends with stellar mass, all three models predict that galaxy sizes decrease with increasing redshift.
From $z=2$ to $z=0$, galaxies with $M_*\approx10^{10}~\msun$ increase in size by $\approx 0.4$~dex for the two SA models and by $\approx 0.2$~dex for \eg. 
For all models, the galaxy sizes are found to be sensitive to the input physics, in particular the stellar feedback \citep[e.g.][]{crain15,lacey15}.

The most notable difference between the models is the decrease in galaxy sizes with increasing stellar mass in the \gl~model, which is not seen for \lgl~and only seen at $z=2$ for \eg, but in a milder form. The decrease in median size for the \gl~galaxies is due to the particular modelling of the self-gravity of discs and how this affects the host dark matter halo (\S\ref{sec:samp}). While \lgl~ignores the self-gravity of baryons and possible contraction of dark matter haloes due to it, \gl~appears to overestimate these effects. The model of the self-gravity of baryons impacts galaxy sizes and this, in turn, affects the evolution of their star formation. Thus, this point is of crucial importance in understanding the evolution of galaxies. The comparison between \eg~and the SA models highlights the need for a better analytic approximation of the effect that baryons have on the distribution of dark matter. 

A further consideration in the SA model modelling of galaxy sizes that can account for some of the differences seen relative to \eg~is the very simple assumptions made to model the angular momentum, in particular for mergers. The evolution of angular momentum, the self-gravity of baryons and its effect on the dark matter are taken into account naturally by gas dynamics in hydro-simulations such an \eg. Although note that hydrodynamical simulations can suffer from angular losses resulting in sizes that are too small if efficient feedback in not implemented \citep{katz92,Navarro93,crain15}. 

In \eg, the decrease in galaxy sizes with increasing stellar mass at $z= 2$ for $M_* \gtrsim 10^{9.7}~\msun$ is not due to an aperture effect. This decline could be due to the highest-mass galaxies at this redshift forming early when densities in the Universe were higher, thus forming compact cores. The comparison to the SA models suggest that the decline in the $M_{*}-r_{50}$ relation at high redshift in \eg~could also be the result of moving from a regime in which the self-gravity of baryons affects the concentration of the dark matter host haloes to a regime in which this effect is erased due to an important contribution from dissipationless mergers \citep{Navarro93,gao04}.

Another interesting difference between the models is the extent of the scatter in the $M_{*}-r_{50}$ relation, which at $z=0$ for $M_*<10^{9.5}~\msun$ ranges from  $\approx 0.2$~dex in \eg, to $\approx 0.4$~dex for the two SA models. Separating the galaxy sample by morphology, we find that the scatter in the relation in \gl~is larger for bulge dominated galaxies than for discs, but for the latter the range still covers $\approx 0.4$~dex. Thus, the difference in the extent of the scatter between the two SA models and \eg~might point to a more fundamental aspect of the modelling related to how the angular momentum of discs is tracked. Exploring this possibility further requires a study of individual galaxies which is beyond the scope of this paper.

\section{Conclusions}\label{sec:conclusions}

We compared global properties of galaxies in the cosmological hydrodynamical simulation \eg~and from two semi-analytic (SA) models of galaxy formation, \lgl~and \gl.  All the models in this comparison include physical prescriptions for the processes considered to be most important for galaxy formation, namely gas cooling, star formation, metal enrichment, feedback from stars and AGN, and, in the case of the SA models, environmental processes (which arise naturally in hydrodynamical simulations). For this comparison, all three models are produced from simulations from the same initial conditions, with the SA models built on merger trees constructed from the \eg~dark matter-only (\egdm) simulation (see Table~\ref{tbl:eagle}). The simulations follow a cubic volume of side 100 comoving Mpc, with 1504$^3$ dark matter particles, and an equivalent number of baryonic particles for \eg, with cosmological parameters set by results from the Planck mission \citep[][see also Table~\ref{tbl:cosmo}]{planck13}. Relative to the published models, only one significant change was implemented, in the case of \gl, a gradual ram-pressure stripping prescription for the hot gas in satellite galaxies replaced the instantaneous one. This change has the primary effect or reducing the fraction of satellite galaxies found to be passive. As a result of this change and the use of the merger trees from the \egdm, both SA models required a modest amount of recalibration to match the observational diagnostics at $z\approx 0$, summarised in Table~\ref{tbl:cali}. But note that no attempt was made to match the SA models to the \eg~results. 

This paper focuses on properties that encapsulate the evolution of typical galaxies with $M_* > 10^8~\msun$.
In particular, we compare stellar masses, including the galaxy stellar mass function (GSMF) and its evolution, mean halo occupation, star formation properties, metallicities and galaxy sizes. By construction, the three GSMFs at $z=0$ are in reasonable agreement, with differences below $0.5$~dex for galaxies with $10^{8.0}<M_*/\msun<10^{11}$. Nevertheless, the evolution with redshift of the GSMF and other properties are not determined by the calibration of the model free parameters (which are set by observations at $z\approx 0$, as summarised in Table~\ref{tbl:cali}). 

The GSMFs at $z\leq 2$ are in reasonable agreement for all three models, with differences in number density below $0.5$~dex for galaxies with $10^{8.0}<M_*/\msun<10^{10.5}$ (Fig.~\ref{fig:smf}). The stellar mass densities differ by $\lesssim 0.3$~dex at $0<z<5$ (Fig.~\ref{fig:sm_vs_z}). At each redshift, star-forming galaxies have been defined using the same cut in specific star formation rate (sSFR) for all the models (\S\ref{sec:sf_def}).  At $z\leq 2$, the median sSFR in the three models agree within $\approx0.4$~dex for star-forming galaxies with $10^{8.0}<M_*/\msun<10^{9.5}$ (Fig.~\ref{fig:ssfr_m}). For all three models the median sSFR closely follows the mass assembly history of the host dark matter halo (Fig.~\ref{fig:ssfr_vs_time}). These similarities indicate that the galaxy populations evolve in a consistent way across all models.

In spite of the overall good agreement, some discrepancies were uncovered, which can guide improvements to future models. These differences and their significance are summarised below.

\begin{itemize}
\item Despite the good agreement found for the global GSMFs, the GSMFs for
  central passive galaxies exhibit clear differences
  (Fig.~\ref{fig:fig5_smf_split}). In both \eg~and \lgl~there is an excess of
  central passive galaxies with $M_*<10^{9.5}\msun$ that is not present in \gl.
  This excess is at least partly due to ejected satellite galaxies, i.e.,
  galaxies that were once close to the centre of a larger halo, thus
  experiencing suppressed star formation due to environmental processes, such as
  ram-pressure stripping of hot gas, but that at later times are identified as
  central galaxies in their own haloes by some merger trees construction
  algorithms, such as those used in \eg~and \lgl. In SA models, only the
  galaxies classified as central will be allowed to accrete new gas, which might
  lead to the formation of new stars, making a difference to their SF histories,
  with respect to galaxies classified as satellites.

\item Several differences suggest that in \lgl~the AGN feedback is not as
  efficient in quenching star formation as in the other two models. This
  difference with respect to \eg~and \gl~drives \lgl~to have a higher stellar
  mass density for star-forming galaxies with $M_*>10^8~\msun$ at $z<1.5$
  (Fig.~\ref{fig:sm_vs_z}), a higher stellar mass-halo mass relation for haloes
  with $M_{\rm halo}\approx 10^{12}~\msun$ at $z=0$ (Fig.~\ref{fig:halo_vs_sm}),
  a higher normalisation of the star formation rate density (SFRD) at $z<1.5$
  (Fig.~\ref{fig:sfrd_vs_sm}) and a higher median sSFR for star-forming galaxies
  with $M_*>10^{10.5}~\msun$ at all redshifts explored (Fig.~\ref{fig:ssfr_m}).
  Note that the exact variations in normalisation depend on the definition of
  star-forming galaxies. In order to decide which model has the most realistic
  modelling of the effect of feedback, a detailed comparison with observations
  is required, while simultaneously increasing the sample of simulated massive
  galaxies where AGN are found to have an impact by considering larger volumes.
  Similar comments are also relevant for establishing the level of realism of
  other modelled physical processes, such as the need for a threshold in gas
  density for star formation to happen.

\item In \gl~there is a higher number of star-forming galaxies with
  $M_*<10^{9.5}\msun$ than in the other two models (Fig.
  \ref{fig:fig5_smf_split} and Fig.~\ref{fig:fraction_vs_sm}). This difference
  is related to \gl~having a weaker stellar feedback or possibly a shorter
  reincorporation timescale for the reheated gas in galaxies.  A comprehensive study of this point will be done in a future paper following individual haloes.

\item Environmental processes are naturally accounted for in hydrodynamic
  simulations. The comparison between observations and variations of \gl~with
  \eg~has reinforced the need to model the stripping of the gas from satellite
  galaxies in a gradual manner and not instantaneously (Fig.
  \ref{fig:fig5_smf_split} and Fig.~\ref{fig:fraction_vs_sm}). 

\item The gas and stellar mass-metallicity relations and their evolution are
  very different among the three models, in particular for low-mass galaxies
  (see Figs.~\ref{fig:starz_vs_sm} and \ref{fig:z_vs_sm}).  While the
  mass-metallicity relations do not evolve significantly in the two SA models,
  there is a clear flattening with time of the relation for \eg, which appears
  to be closer to observations (see Figs.~\ref{fig:z_vs_sm}).  The lack of
  evolution for the two SA models is due to the metal mass following, on
  average, the growth of stellar mass, which is not the case in \eg.  Note,
  however, that the mass-metallicity relations in the main \eg~simulation are
  not converged at the low stellar mass end: higher-resolution \eg~simulations
  result in steeper mass-metallicity relations than those presented here
  \citep{schaye15}. In order to better understand the evolution of the
  mass-metallicity relations, higher-resolution simulations are needed for \eg.
  In the case of the SA models,  winds can be treated in a more realistic way,
  for example by using a dynamical model of supernova feedback following the
  evolution of pressurised bubbles \citep{lagos13}. The treatment of winds has a
  direct impact on the evolution of galaxies hosted by small haloes.  The
  justification of these changes will certainly require further detailed
  investigations of the chemical evolution of gas and stars in both
  hydrodynamical simulation, SA models, and observations.


\item The three models also predict different stellar mass-size relations
  (Fig.~\ref{fig:r50}). The differences between the two SA models stem from
  different approaches to modelling the effect that the self-gravity of baryons
  has on both the baryons themselves and on their host dark matter haloes: while
  \lgl~neglects these effects, \gl~uses an approach that overpredicts them. The
  effect that baryons have on the underlying dark matter is naturally accounted
  for in \eg. One other aspect that affects the stellar mass-size relation is
  the simple assumptions made by the two SA models for modelling the evolution
  of the angular momentum. The scatter in the stellar mass-size relation is a
  factor of $\approx 2$ larger in the two SA models compared to \eg. We have not
  found a satisfactory explanation for this difference. It is likely that this
  difference is related to the way the angular momentum of discs is tracked in
  the SA models. To explore this possibility further, it will be useful to
  follow galaxies hosted by the same halo in both \eg~and the SA models.

\end{itemize}

There is a reasonable agreement between \eg~and the SA models in many instances,
which implies that the SA models are adequately encapsulating many of the physical processes relevant for this study that are naturally solved by hydrodynamical
simulations. The level of agreement also shows that the subgrid physics prescriptions in the
hydrodynamical simulation can now result in galaxy populations that have, at least, a
similar level of realism as SA models. In some instances, an agreement between models could be due to a combination of interplaying model processes. Thus, although such similarities can help improve our understanding of the most likely physical process contributing to the evolution of a given galactic property in hydrodynamical simulation, a detailed comparison with observations is needed to advance towards more realistic models. Nevertheless, this comparison between models has highlighted how different modelling techniques can inform future developments by investigating the similarities and differences in the global model galaxy population.
Further insights can for example be gained by using individual object comparisons, together with observational data.

\subsection*{ACKNOWLEDGEMENTS}

We would like to thank Peder Norberg, James Trayford and Carlton Baugh for
fruitful discussions. VGP acknowledges
support from a European Research Council Starting Grant (DEGAS-259586). 
Qi Guo acknowledges support from the NSFC grant (Nos. 11133003), the Strategic
Priority Research Program The Emergence of Cosmological Structure of the Chinese
Academy of Sciences (No. XDB09000000) and the ``Recruitment Program of Global
Youth Experts'' of China, the NAOC grant (Y434011V01).
CPL is funded by an Australian Research Council Discovery Early Career Researcher Award (DE150100618). CSF acknowledges the ERC Advanced Investigation grant "Cosmiway" (GA 267291). RAC is a Royal Society University Research Fellow. This work was supported by the Science and Technology
Facilities Council (ST/L00075X/1); European Research Council (GA 267291 and GA 259586) and by
the Interuniversity Attraction Poles Programme initiated
by the Belgian Science Policy Oce (AP P7/08 CHARM). This work used the DiRAC Data Centric system at Durham University, operated by the Institute for Computational Cosmology on behalf of the STFC DiRAC HPC Facility (www.dirac.ac.uk). This equipment was funded by BIS National E-infrastructure capital grant ST/K00042X/1, STFC capital grants ST/H008519/1 and ST/K00087X/1, STFC DiRAC Operations grant ST/K003267/1 and Durham University. DiRAC is part of the National E-Infrastructure. We acknowledge PRACE for awarding us access to the Curie machine based in France at TGCC, CEA, Bruyeres-le-Ch\^atel. The data are available from the two lead authors.




\appendix

\section{Halo masses}\label{appendix:halomf}

Halos in both the \eg~simulation and the \lgl~model are identified by combining
the FoF \citep{Davis1985} algorithm with the \subf~code
\citep{Springel2001, Dolag2009}. In both cases, a sphere centred at the minimum
of the gravitational potential of each halo is grown until the mass contained
within a given radius, $R_{200}$, reaches $M_{200}^{\rm crit} =
200\left(4\pi\rho_{\rm crit}(z)R_{200}^3/3\right)$, where $\rho_{\rm
crit}(z)=3H(z)^2/8\pi G$ is the critical density at the redshift of interest. 
The \gl~model, although initially it also uses the FoF haloes and their internal
self-bound substructures as identified by \subf, actually identifies
haloes, the Dhaloes, taking into account the merger tree construction process
\citep[][see also \S\ref{sec:mt}]{jiang14}. The mass of a Dhalo is simply the
sum of the masses of its component subhaloes (note that no particle can belong
to more than one Dhalo). These, by construction, are different from those found
initially in some of the FoF groups. In order to compare the host halo masses of
galaxies predicted by the three considered models, we have estimated the $\mth$
for the identified Dhaloes used by \gl. This is done by using the centres of the
Dhaloes and calculating the mass enclosed in a sphere around this centre with a
mean overdensity equal to 200 times the critical value, i.e. using the same
definition for $\mth$ as in \eg~and \lgl. Although this definition is similar,
the centres of the Dhaloes and those haloes identified with the FoF algorithm
can be different, which will give rise to different halo masses in some cases. 
Throughout this paper we simply refer to the $\mth$ obtained as described above
as $M_{\rm halo}$.

In \gl, the reconstructed $\mth$ can be bigger than the original Dhalo mass by
up to a factor of 4 for haloes hosting small galaxies with stellar masses below
$10^8~\msun$. This happens because the $R_{200}$ of these small haloes actually
encloses mass from their neighbouring haloes, which will be counted towards
their $\mth$. Thus, no galaxy with stellar mass below $10^8~\msun$ will be
considered in the comparison. Note that, in any case, $10^8~\msun$ is
below the resolution limit of the largest \eg~simulation.

Here we comment on the variations obtained when the native Dhalo mass is used
for \gl~haloes. For galaxies with stellar masses above $10^{8}~\msun$ at $z=0$,
an average shift of 0.07~dex is found between the stellar mass-halo mass relation
(see \S\ref{sec:shmass}) for \gl~haloes when using either $\mth$ or the native
Dhalo masses, with $\mth$ being larger at $z=0$, when the differences are the
largest. A maximum shift of 0.13~dex is found at $z=0$. The differences are
smaller at higher redshifts. Similar results to those discussed in
\S\ref{sec:hod}, are found for the \nm~obtained utilizing the Dhalo mass for
\gl~haloes instead of $\mth$. The same is true for the probability density
distribution of satellite galaxies separated according to the mass of their host
halo, discussed in \S\ref{sec:envir}.  


\subsection{The halo mass function}

\begin{figure}
  \includegraphics[width=3.0in]{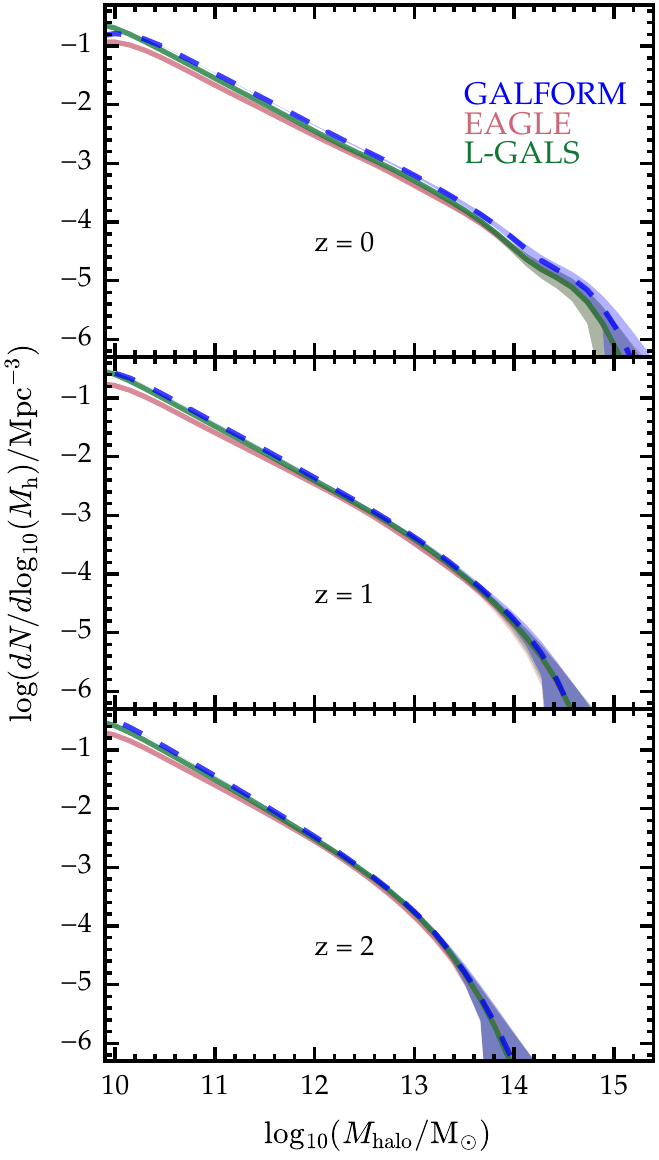}
  \caption{\label{fig:halomf} The halo mass function constructed from the $\mth$
  of the host haloes of all central galaxies in \eg~(red lines), \gl~(black
  dashed lines) and \lgl~(green lines) at $z=0$, $1$ and $2$ from top to bottom.
  The shaded regions show the 1 $\sigma$ range obtained by bootstrapping 200 realisations of the halo mass function. Although the three models are in remarkably good agreement at the massive end, the mass functions from the two SA models are clearly above that for \eg~for haloes with $\lesssim 10^{11.5}~\msun$ due to the early loss of baryons by feedback in \eg.}
\end{figure}

Fig.~\ref{fig:halomf} shows the halo mass function constructed from the $\mth$
of the host haloes of central galaxies in \eg, \gl~and \lgl. At $z=0$, the three
models predict halo mass functions which are in very good agreement above
$10^{12}~\msun$, although, beyond $10^{14}~\msun$ the results are not
statistically significant due to the small numbers of very massive haloes within
the \egdm~simulation. This trend remains true at higher redshifts.

Fig.~\ref{fig:halomf} shows that the halo mass functions
for the two SA models are remarkably similar, with differences ranging from about 5 to
15\%, depending on the halo mass. This difference is likely caused by the
different definitions of central galaxies in the two SA models. 

The halo mass functions predicted by both SA models are above that from \eg~for
haloes below $\approx 10^{13.5}~\msun$, $\approx10^{13}~\msun$ and $\approx
10^{12}~\msun$ at $z=0,\, 1$ and $2$, respectively. At $z=0$, the halo mass
function derived from \gl~is actually above that from \eg~for all the explored
halo masses. Several authors have previously found that the early loss of
baryons due to stellar feedback can reduce the growth rate of those dark matter
haloes that have masses below $10^{13}~\msun$ at $z=0$ \citep[e.g.][]{sawala13}.
\citet{schaller14} found that, in the \eg~simulation, the reduction in halo mass also happens for haloes
hosting galaxies affected by AGN feedback, and that this change is milder than
for those dominated by stellar feedback.

\section{Calibration of the SA models}\label{app_cal}

In this work we make use of two SA models, \lgl~and \gl, based upon the
published versions described in \cite{qi13} and \cite{gp14}, respectively. As
described in detail in \S\ref{sec:models}, the two SA models have been run using
merger trees from the \egdm~ simulation, which assumes a different cosmology and
has a higher mass resolution than the underlying simulation used in these
published models. Moreover, the merger trees from the \egdm\ simulation were
constructed based on 200 snapshots, while those used by \cite{qi13} and
\cite{gp14} were based on only 64. Although the time resolution can affect the
implementation of SA models, our tests with these two flavours of \lgl~ and \gl,
showed the impact to be minimal for the global properties studied here.  In
order to reduce the initial differences between \eg, \lgl~and \gl, the same
Chabrier IMF \citep{cah03} was adopted in \gl, with the consequent modification
in the metal yield and recycled fractions. The choice of stellar population
synthesis (SPS) model was changed from BC99 \citep[an update of][]{bc93} to that
from \citet{cw10} (CW10), which is closer to the SPS from \cite{Bruzual03}
assumed in both \eg~and \lgl. This change of SPS results in negligible
differences in all the studied properties. In order for the passive fractions at
$z=0$ from the three studied models to give a reasonable match to observations,
one additional change was included in the \gl\ model: a gradual ram-pressure
stripping of the hot gas in satellite galaxies. These changes, together with the
change in cosmology and mass resolution of the underlying simulation, resulted
in model luminosity functions and galaxy stellar mass functions at $z=0$ that do
not agree as well with observations as the published models (see
\S\ref{sec:calibration} for more details). Thus, a small adjustment of the model
parameters controlling the feedback was made. The changes with respect to the
published models are summarised in Tables~\ref{tbl:lgal_cal} and
\ref{tbl:galform_cal}. 

\begin{table} 
  \caption{Modifications, besides the cosmology, to the \gl\ model used in this
  work with respect to the published model described in \citet{gp14}. Note that
  $\alpha_{\rm cool}$ is one of the parameters setting the AGN feedback
  efficiency in \gl~\citep{lacey15}.
  }
\begin{center}
\begin{tabular}{|c|c|c|}
\hline
Parameter & \citet{gp14} & This Work \\
\hline
   IMF             &   \citeauthor{kennicutt_imf} & \citeauthor{cah03} \\
   Yield             &  0.021 & 0.02908 \\
   Recycle fraction     & 0.44  & 0.4588 \\ 
   SPS model   &    BC99 & CW10\\
   Stripping of hot gas   &   Instantaneous & Gradual \\
   $\alpha_{\rm cool}$ (AGN feedback)    &    0.60 & 0.52 \\
\hline
\end{tabular}
\end{center}
\label{tbl:galform_cal}
\end{table}

 \begin{table} 
  \caption{Modifications, besides the cosmology, to the \lgl~model used in this work with respect to the published model described in \citet{qi13}. In the \citet{qi13} model, the parameters $\mathit{k}$ and  $\epsilon$ are related to the galaxy feedback, as stated in the table.}
\begin{center}
\begin{tabular}{|c|c|c|}
\hline
Parameter & \citet{qi13} & This Work \\
\hline
   Yield             &  0.03 & 0.02908 \\
   Recycle fraction     & 0.43  & 0.4588 \\ 
   AGN efficiency, $\mathit{k} (10^{-5}~\msun)$   &    $0.7$    &  $3$ \\
   Threshold mass of cold gas reheated \\
   due to star formation, $\epsilon$ & 4.0 & 4.5 \\
\hline
\end{tabular}
\end{center}
\label{tbl:lgal_cal}
\end{table}

\lgl~was calibrated using the GSMF at $z=0$, among other properties (see
\S~\ref{sec:calibration} for more details). The changes in the parameters
detailed in Tables~\ref{tbl:lgal_cal} result in a different GSMF, as shown in
Fig.~\ref{fig:lgal_cal}. This Figure suggests that the SN and AGN feedback are
more efficient in the model used for this work, compared with the published
model in \citet{qi13}.

\begin{figure}
  \includegraphics[width=3.0in]{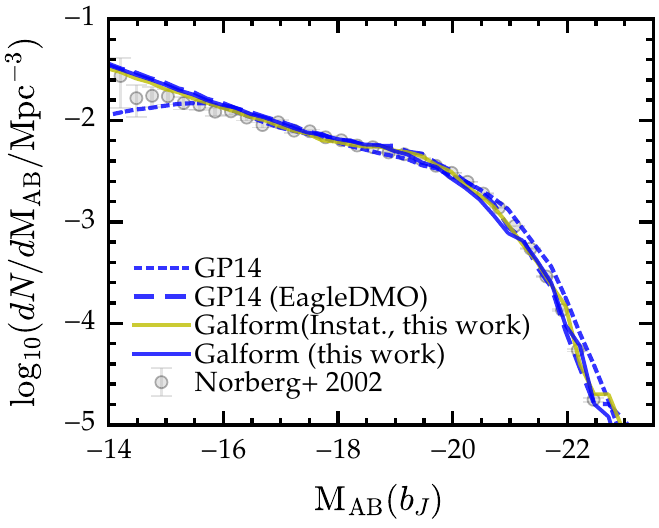}
  \includegraphics[width=3.0in]{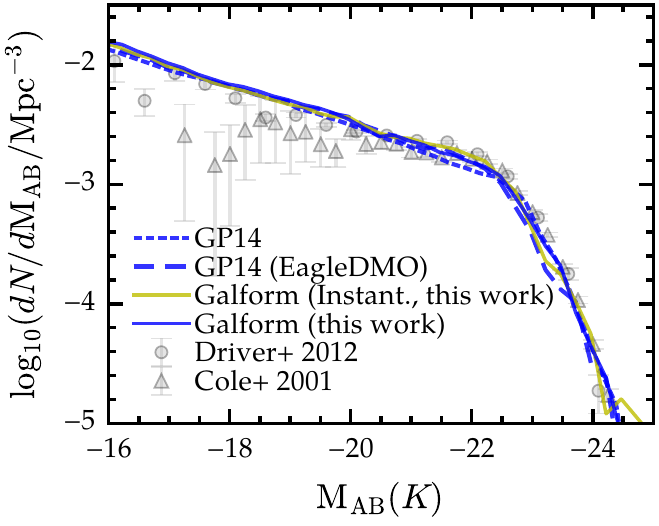}
  \caption{\label{fig:galform_cal} The predicted luminosity functions at
  redshift $z=0$, in the ${\rm b_J}$-band (top panel) and in the K-band (bottom
  panel) which are used by the \gl~model in the calibration against the
  observations. The blue dashed lines are the predictions in \citet{gp14}. The dashed lines are the predictions of GP14 model based on \egdm. The yellow solid lines and blue solid lines are the predictions of the re-calibrated \gl~model with instant
  and gradual ram-pressuring stripping in this work respectively. The
  observational ${\rm b_J}$-band and K-band luminosity functions  are plotted as
  grey points \citep{norberg02} in top panel and 
  grey triangles\citep{cole01}, grey points \citep{driver12} in
  bottom panel respectively. }
\end{figure}

The observed ${\rm b}_{\rm J}$-band and $\rm{K}$-band luminosity functions at $z=0$ are shown in Fig.~\ref{fig:galform_cal} together with those from the \gl~model used for this study both with and instantaneous and a gradual ram-pressure stripping. These are the main observations used to calibrate the \gl~models (see
\S~\ref{sec:calibration} for further details).

Figs.~\ref{fig:galform_cal} and~\ref{fig:lgal_cal} compare the published models, based on merger trees from the MS-W7 simulation (short dashed lines), with exactly the same models but run on the \egdm\ simulation merger trees (long dashed lines). The change in the mass resolution, from the MS-W7 to the \egdm\ simulation, is clear from comparing the two pairs of lines at small masses or luminosities. When the mass resolution is increased in the simulations, more model galaxies with smaller masses or fainter magnitudes are found. This comparison also shows that the combination of the change of cosmology together with the change of mass resolution is model dependent. A detailed study of this point, although interesting, is beyond the scope of this paper.

Fig~\ref{fig:galform_cal} and~\ref{fig:lgal_cal} clearly show that the changes introduced in the SA models result in very small variations of the global properties used for their calibration, at least in the range where there is a large enough number of galaxies within the \egdm\ simulation, as to be statistically meaningful.



\begin{figure}
  \includegraphics[width=3.0in]{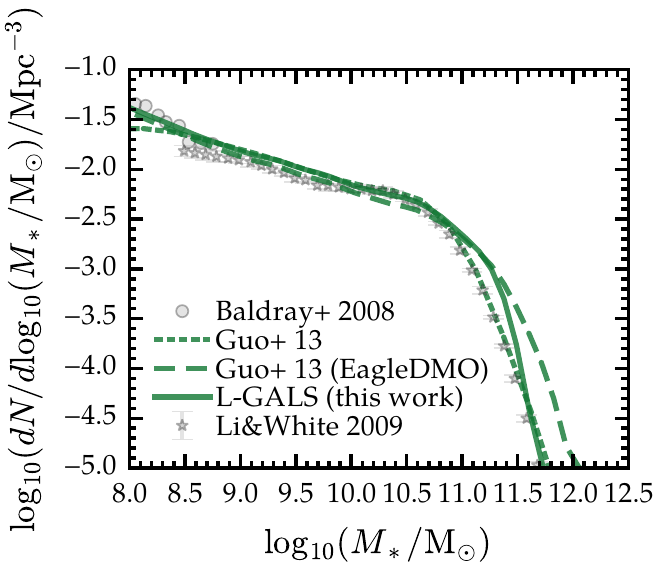}
  \caption{\label{fig:lgal_cal} The predicted GSMF at redshift $z=0$ which are
  used by the \lgl~model in the calibration against the observations. The green
  dotted line are the prediction by \citet{qi13}. The green dashed lines are the
  predictions by \lgl~model with the parameters are the same as in \citet{qi13}
  but based on \egdm. The black solid are the predictions of re-calibrated
  \lgl~model used in this work. The observations used in the calibration are
  plotted as grey points \citep{baldry08} and grey stars \citep{liw09}.}
\end{figure}


\end{document}